\documentclass[12pt]{article}
\usepackage{amsmath, amsthm} 
\usepackage{multirow} 
\usepackage{amsfonts}
\usepackage{amssymb,graphics,psfrag}
\usepackage{array,epsfig,multirow,stmaryrd,graphicx}
\usepackage{comment}
\def\hybrid{\topmargin -20pt    \oddsidemargin 0pt
        \headheight 0pt \headsep 0pt
        \textwidth 6.5in        % US paper
        \textheight 9in         % US paper
        \marginparwidth .875in
        \parskip 5pt plus 1pt   \jot = 1.5ex}
\hybrid
\numberwithin{equation}{section}
\numberwithin{table}{section}
\newcommand{\ds}{\displaystyle}
\newcommand{\cO}{\mathcal{O}}

\newcommand{\cC}{\mathcal{C}}
\newcommand{\cD}{\mathcal{D}}
\newcommand{\cL}{\mathcal{L}}

\newcommand{\cW}{\mathcal{W}}

\newcommand{\cA}{\mathcal{A}}

\newcommand{\cF}{\mathcal{F}}

\newcommand{\MM}{\mathcal{M}}
\newcommand{\cM}{\mathcal M}

\newcommand{\cref}{{\bf [check ref]}}

\newcommand{\CD}{{\cal D}}

\newcommand{\CZ}{{\cal Z}}
\newcommand{\dd}{{\rm d}} 

\def\Z{{\mathbb Z}}
\def\R{{\mathbb R}}
\def\C{{\mathbb C}}
\def\P{{\mathbb P}}

\newcommand{\der}{\partial}

\def\blfootnote{\xdef\@thefnmark{}\@footnotetext} 
\long\def\symbolfootnote[#1]#2{\begingroup%
\def\thefootnote{\fnsymbol{footnote}}\footnote[#1]{#2}\endgroup}

\begin{document}

%%%%%%%%%%%%%%%%%%%%%%%%%%%%%%%%%%%%%%%%%%%%%%%%%%%%%%%%%%%%%%%%%%%%%%%%%%%%%%%%%

\baselineskip=18pt

\begin{titlepage}
\begin{flushright}
\parbox[t]{1.8in}{
BONN-TH-2009-05}
\end{flushright}

\begin{center}

\vspace*{ 1.2cm}

{\large \bf \Large Solving the Topological String on K3 Fibrations}

\vskip 1.2cm

\begin{center}
 \bf{Babak Haghighat\footnote{babak@th.physik.uni-bonn.de}, Albrecht Klemm\footnote{aklemm@th.physik.uni-bonn.de}}
\end{center}
\vskip 0.2cm

{\em Physikalisches Institut der Universit\"at Bonn \\[.2cm]
and\\[.1cm]
Bethe Center for Theoretical Physics\\[.1cm] 
Nu\ss allee 12, 53115 Bonn, Germany}
\vspace*{.1cm}

\end{center}

\vskip 0.2cm

\begin{center} {\bf ABSTRACT } \end{center}
We present solutions of the holomorphic anomaly equations for compact two-parameter Calabi-Yau manifolds 
which are hypersurfaces in weighted projective space. In particular we focus on K3-fibrations where due 
to heterotic type II duality the topological invariants in the fibre direction are encoded in certain
modular forms. The formalism employed provides holomorphic expansions of topological string amplitudes
everywhere in moduli space.

\end{titlepage}

%%%%%%%%%%%%%%%%%%%%%%%%%%%%%%%%%%%%%%%%%%%%%%%%%%%%%%%%%%%%%%%%%%%%%%%%%%%%%%%%%%%%%%%%%%%%%
%%%%%%%%%%%%%%%%%%%%%%%%%%%%%%%%%%%%%%%%%%%%%%%%%%%%%%%%%%%%%%%%%%%%%%%%%%%%%%%%%%%%%%%%%%%%%

\newpage

\tableofcontents

\begin{flushleft}
  \underline{\hspace{16.5cm}}
\end{flushleft}

\section{Introduction}

Recent progress in the computation of topological string amplitudes to higher genus on compact
Calabi-Yau manifolds \cite{HKQ} \cite{HoKo} \cite{GrCY} raises the question of integrability of the 
closed topological string on these backgrounds. The method employed in the above papers uses the
polynomial structure of topological partition functions first found by Yamaguchi and Yau in \cite{YamYau}.
This method was generalized in \cite{GKMW} \cite{alimlaenge} and was applied to local Calabi-Yau manifolds 
in \cite{integrability} where it was found that the gap condition of
\cite{HKQ} together with regularity at other points in the moduli space is enough to fix
the holomorphic ambiguities completely. This makes the topological string
integrable on non-compact Calabi-Yau manifolds whose mirror geometry is
encoded by a family of Riemann surfaces $\Sigma_g$ and a meromorphic
differential $\lambda$. Indeed the successful direct integration in the case of non-compact models can be traced back to well known transformation properties of the topological string amplitudes w.r.t to the modular
group of $\Sigma_g$, which is a finite index subgroup of $Sp(2g,\Z)$. Moreover
it can be established~\cite{Marino:2006hs,Bouchard:2007ys} that the theory becomes equivalent to a matrix model 
whose eigenvalue dynamics and correlators can be completely fixed by its spectral
curve and the meromorphic differential defining the filling fraction. These
data are precisely identified with the Riemann surface $\Sigma_g$ and
$\lambda$ respectively~\cite{Marino:2006hs,Bouchard:2007ys}. 
However, in the case of compact Calabi-Yau manifolds the situation is much more involved as the modular
groups are not well understood and a classification of automorphic forms on the moduli spaces 
is still lacking. Also, there is no description in terms of a matrix model available, yet.

In this paper we construct solutions to the holomorphic anomaly equations for
regular K3-fibrations with two moduli, which are realized as hypersurfaces in toric 
ambient spaces. We focus in particular on two models whose moduli spaces have
been explored in detail~\cite{CandelasMirrorII}. These moduli spaces have
identical types of boundary divisors, which have been interpreted physically
in the context of heterotic/type II duality. Apart from the large radius point one has the weak coupling
divisor, the strong coupling divisor, a Seiberg-Witten divisor, the generic
conifold locus and the Gepner point where a conformal field theory 
description becomes available. The strong and weak coupling divisors owe 
their names to their interpretation in the dual heterotic models where the 
size of the $\P^1$-base of the K3-fibration is mapped to the value of the dilaton. One of these 
models is a degree $12$ hypersurface in $\P_4^{(1,1,2,2,6)}$ which appeared in the work of Kachru and
Vafa \cite{KachruVafa} where its heterotic dual, namely the $ST$ model, was found by identifying
the prepotentials of both theories. 
Although there has not yet been found a heterotic dual for the other model which is a degree
$8$ hypersurface in $\P_4^{(1,1,2,2,2)}$ the prepotentials as well as the higher genus free energies
$F_g$ for both models have been reconstructed completely in the fiber direction by a duality inspired 
analysis \cite{KlemmScheidegger}. We use these results as well as boundary information
from other divisors in moduli space to find global expressions for the $F_g$'s. In particular we find
that the gap condition of \cite{HKQ} holds at the conifold locus and is enough to fix all
ambiguities related to the conifold discriminant. Furthermore, at the strong coupling divisor we observe a
gap for genus $2$ and $3$. However, this gap goes away at genus $4$ which is a hint that there are
nontrivial interactions between the light hyper- and vectormultiplets around this divisor.
Last but not least regularity at the Gepner point together with the K3 fibre results of \cite{KlemmScheidegger}
provide enough conditions to solve for all parameters in the holomorphic ambiguity up to genus 4. 

The paper is organized as follows. In section \ref{completeintersection} we review the construction of 
three dimensional Calabi-Yau hypersurfaces in toric ambient spaces. The symmetries of the ambient space are used
in section \ref{PFeq} to derive Picard-Fuchs equations for the threefold which will be used in subsequent sections
to compute periods and mirror map for the relevant Calabi-Yau. Section \ref{holanomalyeqs} reviews as a first step
the results of Berchadsky, Ceccoti, Ooguri and Vafa on the holomorphic anomaly equations and as a second
step outlines a method of solution relying on the differential ring structure generated by the propagators.
Special emphasis is put on the parametrization of the holomorphic ambiguities. We then move on to a description of
the moduli space of K3 fibrations where we give a detailed presentation of the physically important
boundary divisors. The final section is devoted to the concrete solution of the two models and we display our
results for the topological free energies around several important points in moduli space.

%%%%%%%%%%%%%%%%%%%%%%%%%%%%%%%%%%%%%%%%%%%%%%%%%%%%%%%%%%%%%%%%%%%%%%%%%%%%%%%%%%%%%%%%%%%%%
%%%%%%%%%%%%%%%%%%%%%%%%%%%%%%%%%%%%%%%%%%%%%%%%%%%%%%%%%%%%%%%%%%%%%%%%%%%%%%%%%%%%%%%%%%%%%
\section{Topological Strings on compact Calabi-Yau}
\label{topstronccy}
This section reviews necessary facts about the topological string on compact 
Calabi-Yau manifolds defined as hypersurfaces in toric ambient spaces.
First of all we describe the mirror construction for such Calabi-Yau
manifolds\cite{batyrev}. We then discuss the holomorphic anomaly
equations~\cite{BCOV} and describe the method of direct integration for 
solving them. The construction of~\cite{batyrev} provides us with 
a large class of mirror pairs. Note however that once (\ref{ringofpicardfuchs}) 
and (\ref{largeradiusprepotential}) are given for any construction of 
Calabi-Yau manifold mirror pairs the integration of higher genus 
topological string amplitudes proceeds from general principles 
of Calabi-Yau 3-folds: Unobstructedness of the moduli space, 
special geometry, modularity and boundary conditions, which rely 
on general properties of $N=2$ supergravity effective actions in 4d. 
One particular property of our two examples is however the existence 
of an exact conformal field theory description by an orbifold of 
a tensor product of the minimal 2d superconformal field theories 
at the so called Gepner point in the complex moduli space.                   

\subsection{Calabi-Yau hypersurfaces in toric varieties}
\label{completeintersection}
Toric ambient spaces $\P_{\Sigma}$ of complex dimension $d$ are described through the quotient

\begin{equation}
  \P_{\Sigma} = (\C^n - Z)/G,
\label{psigma}
\end{equation}
where $G \cong (\C^*)^h$ with $h=n-d$ and one has to exclude an exceptional set $Z \subset \C^n$ to obtain a well-behaved
quotient. The $h$ independent $\C^*$ identifications arise as follows. $\P_{\Sigma}$ is defined in terms
of a fan $\Sigma$, which is a collection of rational polyhedral cones $\sigma \in \Sigma$ containing
all faces and intersections of its elements \cite{Fulton,Cox}. The cones are spanned by vectors
which are sitting in a $d$-dimensional integral lattice $\Gamma^*$ and $\P_{\Sigma}$ is compact if the support
of $\Sigma$ covers all of the real extension $\Gamma^*_{\R} = \Gamma^* \otimes \R$ of the lattice $\Gamma^*$. We will concentrate
on the case where $\Sigma$ consists of the cones over the faces of an integral
polyhedron $\Delta^* \subset \Gamma^*_{\R}$, which contains the origin $v_0=(0,\ldots,0)$.
In toric geometry $l$-dimensional cones of $\Delta^*$ represent codimension $l$ subvarieties of $\P_{\Sigma}$.
Now, let $\Sigma(1)$ denote the set of one-dimensional cones with primitive generators $v_i, i=1,\cdots, n$.
One finds that there are $h$ $n$-vectors $Q_i^a\in \Z, a=1,\cdots,h$, called charge vectors,
satisfying the linear relations $\sum_{i=1}^n Q_i^a v_i = 0$ among 
the primitive lattice vectors $v_i$. This defines an action of the group $G$ on the homogeneous coordinates $x_i \in \C$ 
as follows: $x_i \mapsto \mu_a^{Q_i^a} x_i$ with $\mu_a \in \C^*$.

Anti-canonical hypersurfaces in $\P_{\Sigma}$ are given by sections of the anti-canonical bundle $\cO_{\P_{\Sigma}}(\sum_{v_i \in \Sigma(1)} D_i)$,
where $D_i$ is the corresponding divisor to $v_i \in \Sigma(1)$. In order for these to be Calabi-Yau a further condition must
be satisfied. $\P_{\Sigma}$ will usually have singularities which have to be blown up and the criterion for the canonical bundle
to extend to a bundle of the blow-up is that the singularities are of Gorenstein type. Once we also require $\P_{\Sigma}$ to be Fano,
i.e. that the anti-canonical bundle is positive, the above hypersurfaces will define Calabi-Yau hypersurfaces $M \subset \P_{\Sigma}$.

Let us now pass over to the description of mirror symmetry for these
Calabi-Yau hypersurfaces. We denote by $\Gamma$ the dual lattice to $\Gamma^*$, $\Gamma_{\R}$
the real extension and by $\langle \cdot, \cdot \rangle$ the canonical pairing
between the dual vector spaces and define the dual polyhedron $\Delta$ as 
\begin{equation}
  \Delta = \{m\in \Gamma_{\R} | \langle n,m\rangle \geq -1 \textrm{~~for all~~} n \in \Delta^*\}.
\label{reflexive}
\end{equation}
If all vertices of $\Delta$ belong to $\Gamma$ and $\Delta$ contains the origin, then $\Delta$ is 
again an integral polyhedron and both $\Delta^*$ as well as $\Delta$ are called reflexive. 
Note that this implies that in both $\Delta^*$ and $\Delta$ the origin is the only 
interior point. In \cite{batyrev} Batyrev showed that $\Delta$ is reflexive if and only if
the corresponding toric variety, denoted by $\P_{\Sigma^*}$, is Gorenstein and
Fano. This opens up the way for the construction of the mirror Calabi-Yau manifold as a hypersurface in $\P_{\Sigma^*}$, where 
$\Sigma^*$ is the fan over the faces of $\Delta$. The construction uses the fact
that the toric variety corresponding to a fan $\Sigma$ can be defined alternatively through the polyhedron $\Delta$ as an embedding
$\P_{\Sigma}=\P_{\Sigma(\Delta)} \hookrightarrow \P^k$ with $k=|\Delta\cap \Gamma|-1$ using the linear relations among the vertices of $\Delta$.
The same applies to the toric variety corresponding to $\Sigma^*$, where now $\P_{\Sigma^*}=\P_{\Sigma(\Delta^*)} \hookrightarrow \P^{k'}$ with
$k'=|\Delta^* \cap \Gamma^*|-1$. Batyrev showed that the mirror Calabi-Yau manifold $W$ is given by the anti-canonical 
hypersurface in $\P_{\Sigma^*}$. The Hodge numbers can be computed through methods of toric geometry 
and one obtains \cite{batyrev}:
\begin{eqnarray} \label{hodge}
  h^{1,1}   & = & l(\Delta^*) - d - 1 - \sum_{\gamma^*} l^*(\gamma^*) + \sum_{\Theta^*} l^*(\Theta^*) l^*(\hat{\Theta}^*) \\
  h^{d-2,1} & = & l(\Delta) - d - 1 - \sum_{\gamma} l^*(\gamma) + \sum_{\Theta} l^*(\Theta) l^*(\hat{\Theta})\ .
\end{eqnarray} 
Here $\gamma^*$ ($\gamma$) refers to codimension 1 faces of $\Delta^*$ ($\Delta$) and  $\Theta^*$ ($\Theta$) 
refers to codimension 2 faces of $\Delta^*$ ($\Delta$). By $\hat{\Theta}^*$ we denote the face of $\Delta$, which is 
dual to $\Theta^*$ in $\Delta^*$ and vice versa. If $F$ is a facet of the polytop $\Delta$ or $\Delta^*$ then  $l(F)$ denotes the set of 
all integral points on $F$, while $l^*(F)$ denotes only the interior integral points, i.e. those which do not lie in 
codimension one facets of $F$.

In the following we will describe the case $h=1$ and $d=4$, i.e. $3$-dimensional hypersurfaces in weighted projective space,
as this is the relevant construction for our particular models. We also assume that one weight is $1$ and all weights 
divide the degree $D$ of the anticanonical hypersurface and denote the weight vector by $(Q_1,Q_2,Q_3,Q_4,1)$.  Then 
$\Delta^*$ spanned by the vertices 
\begin{displaymath}
  \begin{array}{cccc}
    v_1=(1,0,0,0), & v_2 = (0,1,0,0), & v_3=(0,0,1,0), & v_4 = (0,0,0,1),\\
    \multicolumn{4}{c}{v_5= (-Q_1,-Q_2,-Q_3,-Q_4)}\\
  \end{array}
\end{displaymath}
is an reflexive polyhedron and the charge vector is identified with the weight vector. It is 
convenient to consider also the extended vertices $\bar v_i=(1,v_i)$. The linear relation 
between the extended vertices $\sum_{i=1}^5 Q_i \bar v_i = D \bar v_0$ reproduces the  Calabi-Yau 
condition for $c_1(TM)=0$ namely $D=\sum_{i} Q_i$, where $d$ is the degree of the hypersurface.

According to the above description this on the one hand defines the toric variety $\P_{\Sigma}$ as the weighted projective space 
$\P_d^{(\vec{Q})}$ with the family of Calabi-Yau hypersurfaces $M$ given by generic degree $D$ homogeneous polynomials and
we write $M = \P_d^{(\vec{Q})}[D] \subset \P_{\Sigma}$.
On the other hand the linear dependence of the charge vectors leads to the identification
\begin{equation}
  \P_{\Sigma^*}=\P_{\Sigma(\Delta^*)} \equiv \mathbf{H}^5(\vec{Q}) := \{(U_0,U_1,U_2,U_3,U_4,U_5) \in \P_5 | \prod_{i=1}^5 U_i^{Q_i} = U_0^D\}.
\end{equation}

One can now consider the embedding map $\phi: \P_4^{(\vec{Q})} \rightarrow \mathbf{H}^5(\vec{Q})$ given by
\begin{equation} \label{orbmap}
  [y_1,y_2,y_3,y_4,y_5] \mapsto [y_1 y_2 y_3 y_4 y_5, y_1^{D/Q_1},y_2^{D/Q_2},y_3^{D/Q_3},y_4^{D/Q_4},y_5^{D/Q_5}],
\end{equation}
which defines the isomorphism $\P_{\Sigma^*} \cong \P_4^{(\vec{Q})}/\textrm{Ker}\, \phi$. Anti-canonical hypersurfaces
in $\P_{\Sigma^*}$ are defined through  expressions linear in the $U_i$ which in turn can be expressed as 
monomials in the $y_i$ through equation (\ref{orbmap}). Resolution of the singularities arising from $\P_{\Sigma^*}$ then
gives the family of mirror Calabi-Yau hypersurfaces $W \subset \P_{\Sigma^*}$.   

\subsection{Picard-Fuchs equations and the B-model}
\label{PFeq}

We want to analyze the periods of the mirror Calabi-Yau. The mirror is given by sections of the anti-canonical bundle of $\P_{\Sigma^*}$.
These can be identified with the Laurent polynomials
\begin{equation}
  f = \sum_i a_i Y^{m_i}, \quad Y^{m_i} = Y_1^{m_i^1} Y_2^{m_i^2} \cdots Y_d^{m_i^d},
\end{equation}
where $(Y_1, \cdots, Y_d)$ are coordinates for the torus $T \subset \P_{\Sigma^*}$ and $m_i$ are points of $\Delta^* \cap \Gamma^*$ which do not lie
in the interior of codimension one faces of $\Delta^*$. The Griffiths construction \cite{Griffiths} then gives the
following set of Periods:
\begin{equation} \label{periods}
  \Pi_i(a) = \int_{\gamma_i}\Omega= \int_{\gamma_i} \frac{1}{f(a,Y)} \prod_{j=1}^d \frac{d Y_j}{Y_j},
\end{equation}                                               
with $\gamma_i \in H_d((\C^*)^d \backslash Z_f)$. Here $Z_f$ is the vanishing locus of the polynomial $f$.
These periods satisfy a set of differential equations which are called the GKZ system \cite{GKZ}.
In order to obtain them, we extend the vectors $v_i$ from above to the $\R^{d+1}$ dimensional vectors $\bar{v}_i = (1,v_i)$ forming
the set $A=\{\bar{v}_0,\cdots,\bar{v}_n\}$. Assuming that these integral points span $\Z^{d+1}$ we obtain $h=n-d$ linear dependencies
described by the lattice
\begin{equation}
  \Lambda = \{(l^{(k)}_0,\cdots,l^{(k)}_n) \in \Z^{n+1} | \sum_{i=0}^n l^{(k)}_i \bar{v}_i = 0, \ k=1,\ldots,h \}.
\end{equation}
Now we are ready to write down the differential operators which annihilate the periods (\ref{periods}):
\begin{equation} \label{PFop}
  \CD_k = \prod_{l^{(k)}_i>0} \left(\frac{\der}{\der a_i}\right)^{l^{(k)}_i} - \prod_{l^{(k)}_i<0} 
\left(\frac{\der}{\der a_i}\right)^{-l^{(k)}_i},
\end{equation}
for each element $l^{(k)}$ of $\Lambda$ and 
\begin{equation} \label{scaleinv}
  \CZ_j = \sum_{i=0}^n \bar{v}_{i,j}a_i \frac{\der}{\der a_i} - \beta_j,
\end{equation}
where $\beta \in \R^{d+1}$ and $\bar{v}_{i,j}$ represent the $j$-th component of the vector $\bar{v}_i \in \R^{d+1}$.
One can show that equation (\ref{scaleinv}) defines invariance under the rescalings $a_i \mapsto \lambda_j^{m_{i,j}} a_i$
and $f \mapsto c \cdot f$ for $\lambda$, $c$ $\in \C^*$. Therefore we define the invariant variable 
\begin{equation} \label{zmoduli}
  z_j = (-1)^{l_0^{(k)}} \prod_i a_i^{l^{(j)}_{i}},
\end{equation}
which transforms (\ref{PFop}) to a generalized system of hypergeometric equations
\begin{equation}
  \CD_k \Pi_i(z_1, \cdots, z_{|\Lambda|}) = 0
\label{ringofpicardfuchs}  
\end{equation}
for each $l^{(k)} \in \Lambda$.

In general, these differential equations forming the so called $\cA$-system, contain the periods among their solutions, but there
will be also other solutions. The set of Picard-Fuchs equations which vanish only on the periods can be obtained by factoring the
above equations. Then the resulting lower order operator is Picard-Fuchs once it annihilates all periods.

For a general set of Picard-Fuchs equations, the solution space has dimension $h_{3}(W)$ and one obtains the following set of periods 
\cite{CIMS}:

\begin{equation} \label{Wp}
  \Pi(z) = \left(\begin{array}{c} 
                    \omega_0(z,\rho)|_{\rho=0}\\
                    D_i^{(1)} \omega_0(z,\rho)|_{\rho=0}\\
                    D_i^{(2)} \omega_0(z,\rho)|_{\rho=0}\\
                    D^{(3)}   \omega_0(z,\rho)|_{\rho=0}\\
                  \end{array}
            \right).
\end{equation}

Here, $i$ runs from $1$ to $h_{21}(W)$, where $h_{21}(W)$ is the number of moduli. Furthermore we have the following definitions:
\begin{equation}
  \omega_0(z,\rho) = \sum_{n_i \geq 0} c(n + \rho) z^{n + \rho}, 
\end{equation}
\begin{equation}
  D_i^{(1)} := \der_{\rho_i}, ~ D_i^{(2)} := \frac{1}{2} \kappa_{ijk} \der_{\rho_j} \der_{\rho_k},
  ~ D^{(3)} := -\frac{1}{6} \kappa_{ijk} \der_{\rho_i} \der_{\rho_j} \der_{\rho_k},  
\end{equation}

where $\kappa_{ijk}$ are the classical intersection numbers of the Calabi-Yau $M$ and $c(n+\rho)$ is 
defined by
\begin{equation}
c(n+\rho)=\frac{\Gamma\left(\sum_{k=1}^h l_0^{(k)}(n_k+\rho_k)+1\right)}{\prod_{i=1}^n \Gamma\left(\sum_{k=1}^h l_i^{(k)}(n_k+\rho_k)+1\right)}\ .
\end{equation}

 The periods (\ref{Wp}) describe complex structure
deformations of the manifold $W$ and can be written in terms of homogeneous special coordinates $(X^0, X^i)$ as
$(X^0, X^i, (\der \cF/ \der X^i), (\der \cF / \der X^0))$. On the other hand, the period vector $\Pi(t) = (1, t^i, \der_i F, 2F - t^i \der_i F)$
encodes K\"ahler deformations of the Calabi-Yau $M$ with K\"ahler parameter $t^i$. Here $F$ is the prepotential for the K\"ahler side and
admits the formal large radius expansion
\begin{equation}
  F = \frac{1}{6} \kappa_{ijk} t^i t^j t^k + \frac{1}{2} a_{ij} t^i t^j + b_i
  t^i + \frac{1}{2} c + F_{\textrm{inst}}.
\label{largeradiusprepotential} 
\end{equation}
These two period vectors are related around $\textrm{Im}(t^i) \rightarrow \infty$ through $\Pi(z) = X^0 \Pi(t)$ with
the choice 
\begin{equation} \label{defMirrorMap}
  t^i(z) = \frac{\omega_i(z)}{\omega_0(z)},~ \omega_i(z) := D^{(1)}_i \omega_0(z,\rho)|_{\rho=0}\ .
\end{equation}
From the periods we can calculate the triple couplings
\begin{equation}\label{yuk}
C_{ijk}=\int_{W} \Omega\wedge \der_i\der_j \der_k \Omega=D_i D_j D_k {\cal  F} \ .
\end{equation}
Note that the covariant derivatives w.r.t. to the Weil-Petersen metric and the K\"ahler connection
become $\der_{t_i}$  in the coordinates (\ref{defMirrorMap}). This justifies the name flat 
coordinates for the $t_i$.

\subsection{Solving the holomorphic anomaly equations}
\label{holanomalyeqs}

In this section we want to outline a method of solution of the B-model higher genus amplitudes using the holomorphic 
anomaly equations~\cite{BCOV}, the modular properties of the $F_g$~\cite{YamYau,alimlaenge,Grimm:2007tm} 
and boundary conditions in particular the gap conditions of \cite{HKQ}.

\subsubsection{The holomorphic anomaly  equations}
\label{holomorphicanmalyequation} 

Topological string theory describes the coupling of topological sigma models to worldsheet gravity. In particular we want to concentrate
on the case of the topological B-model which couples the B-twisted sigma model to worldsheet gravity. Here the genus $g$ free energies 
$\cF^{(g)}$ are sections of the line bundle $\cL$, i.e. $\cF^{(g)} \in \cL^{2-2g}$, where $\cL$ is the line bundle of holomorphic 3-forms over the 
moduli space $\cM$ of Calabi-Yau manifolds. From the point of view of the topological field theory $\cL$ is formed by the charge $(0,0)$ 
subspace of chiral fields. For genus $g > 0$ the $\cF^{(g)}$'s suffer from a holomorphic anomaly first calculated in \cite{BCOV1}\cite{BCOV}. 

For $g=1$ this anomaly takes the form \cite{BCOV1}
\begin{equation} \label{defF1}
  \bar{\der}_{\bar{k}} \der_m \cF^{(1)} = \frac{1}{2} \bar{C}^{ij}_{\bar{k}} C_{mij} - (\frac{\chi}{24}-1)G_{\bar{k} m},
\end{equation}
where
$\bar{C}^{kl}_{\bar{i}} = e^{2K} G^{k \bar{k}} G^{l \bar{l}} \bar{C}_{\bar{i}\bar{k}\bar{l}}$ and $C_{ijl}$ 
are holomorphic Yukawa couplings (\ref{yuk}) which transform as sections of $\textrm{Sym}^3(T\cM) \times \cL^{-2}$. 
Eq (\ref{defF1}) can be integrated to give \footnote{In the following we denote 
the non-holomorphic quantities by calligraphic characters $\cF^{(g)}$ and the holomorphic 
limits by straight characters $F^g_p$, with a label $p$ of the patch, 
where the limit is taken.}
\begin{equation} \label{defF1b}
  \cF^{(1)} = \frac{1}{2}\log\left[\textrm{exp}\left[K(3+h^{1,1}-\frac{\chi}{12})\right]\textrm{det}G^{-1}_{i\bar{j}}|f_1|^2\right].
\end{equation}

Here $K$ is the real K\"ahler potential and one has $\textrm{exp}(K) \sim X^0$ in the holomorphic limit, while $G_{i\bar{j}}$ is the
K\"ahler metric on the complex structure moduli space and its holomorphic limit is given by 
\begin{equation}
  G_{i \bar{j}} \rightarrow \frac{dt^i}{dz_j}.
\end{equation} 

$f_1$ is the holomorphic ambiguity arising from the integration and can be written in terms of the discriminant loci of the Calabi-Yau
moduli space, i.e. $f = \prod_i \Delta_i^{a_i} \prod_{i=1}^{h^{2,1}} z_i^{b_i}$. All free parameters $a_i$, $b_i$ are obtained through the limiting
behavior of $\cF^{(1)}$ near singularities. Canonical boundary conditions are given by the limit 
\begin{equation}
  \lim_{z_i \rightarrow 0} \cF^{(1)} = - \frac{1}{24} \sum_{i=1}^{h^{2,1}} \log(z_i) \int_M c_2 J_i
\end{equation}
as well as by the universal behavior at conifold singularities $a_{\textrm{con}} = - \frac{1}{12}$.

For higher genus ($g \geq 2$) the $\cF^{(g)}$ satisfy recursive holomorphic anomaly equations \cite{BCOV}
\begin{equation} \label{holan}
  \bar{\der}_{\bar{i}} \cF^{(g)} = \frac{1}{2} \bar{C}^{jk}_{\bar{i}} \left(D_j D_k \cF^{(g-1)} + \sum_{r=1}^{g-1} D_j \cF^{(g-r)} D_k \cF^{(r)}\right), 
  \quad (g>1)
\end{equation}
which contain the covariant derivative with respect to the metric on the moduli space and the line bundle $\cL$.

\subsubsection{Direct Integration}
\label{directintegration} 

The method of direct integration relies on four key properties. The first is the fact that the $F_g$ fulfill the
holomorphic anomaly equations.  The second is the fact  that the $F_g$ are modular invariant 
under the monodromy group $\Gamma$ of the Calabi-Yau target space, which is a  subgroup of ${\rm Sp}(h_3,\mathbb{Z})$, 
and can be built from a finite polynomial ring of modular objects. In the large phase space these objects 
can be identified directly with modular forms under $\Gamma$~\cite{Grimm:2007tm}, while the modular generators 
that appear below are obtained after a projection to the small phase space. The third important ingredient is the 
existence of a canonical antiholomorphic extension of the ring of modular forms to a ring of almost holomorphic 
forms, with the property that the appropriate covariant derivatives closes on the almost holomorphic ring and that the 
antiholomorphic derivative in the holomorphic anomaly equation can be replaced by a derivative w.r.t the 
antiholomorphic generators of the almost holomorphic ring. The integration of the polynomials $F_g$ w.r.t. to the 
antiholomorphic generators leaves a holomorphic modular ambiguity, which are finitely generated over the 
smaller holomorphic ring. The final ingredients are  physical boundary 
conditions at the discriminant components of the Calabi-Yau space, which determine 
the coefficients of the holomorphic modular ambiguity and allow only for a restricted class 
of modular objects in the rings, comparable to requiring restricted cusp behaviour 
for modular forms of $\Gamma_0=Sl(2,\mathbb{Z})$.

Indeed  the comparison to the classical theory of $\Gamma_0$ modular forms of elliptic curves~\cite{zagier} is very 
instructive. The ring of modular forms ${\cal M}_*[E_4,E_6]$ is here generated by the Eisenstein series $E_4$ and $E_6$. The 
covariant derivative is the Mass derivative acting on weight $k$ modular forms by
$D_k=\left(\frac{{\rm d}}{2 \pi i {\rm d} \tau }-\frac{k}{4 \pi {\rm Im}(\tau)}\right)$. 
It does not close on  ${\cal M}_*[E_4,E_6]$, but on the ring of almost holomorphic functions 
${\cal M}^{!}[\hat E_2, E_4,E_6]$, where $\hat E_2$ is the an holomorphic extension of the second 
Eisenstein series $\hat E_2=E_2-\frac{3}{\pi {\rm Im}(\tau)}$. The latter plays the role 
of the anholomorphic propagators in the formalism of~\cite{BCOV}. Moreover a modular form w.r.t. $\Gamma_0$ of  weight $k$ 
fulfills a linear differential equation in the $J$-function of order $k+1$. This the analog of the 
Picard-Fuchs equation and even if we know little about the modular objects of the Calabi-Yau group 
$\Gamma$ it is possible to reconstruct them from the solutions of the Picard-Fuchs system. The 
totally invariant complex parameters ${\underline z}$ on the moduli space play here the r\^ole 
of the $J$-function. It should be noted that this is more than a formal analogy, because in certain 
local limits the formalism of the global Calabi-Yau space reduces to the one of a family of elliptic
surfaces.                        

For the Calabi-Yau case the method of direct integration was developed in the  work of \cite{YamYau} 
for the one parameter models and extended in the work~\cite{alimlaenge} to the multimoduli case. 

We follow the latter one and note that the construction and the properties of the anholomorphic 
objects rely crucially on special geometry relation~\cite{BCOV}
\begin{equation}
  \bar{\der}_{\bar{i}} \Gamma^k_{ij} = \delta^k_i G_{j \bar{i}} + \delta^k_j G_{i\bar{i}} - C_{ijl}\bar{C}^{kl}_{\bar{i}},
\end{equation}
from which one can show \cite{alimlaenge,YamYau}  
\begin{eqnarray} \label{trunc}
  D_i S^{jk} & = & \delta^j_i S^k + \delta^k_i S^j - C_{imn} S^{mj} S^{nk} + h^{jk}_i, \nonumber\\
  D_i S^j   & = & 2 \delta^j_i S - C_{imn} S^m S^{nj} + h^{jk}_i K_k + h^j_i, \nonumber \\
  D_i S     & = & -\frac{1}{2} C_{imn} S^m S^n + \frac{1}{2} h^{mn}_i K_m K_n + h^j_i K_j + h_i, \nonumber \\
  D_i K_j   & = & -K_i K_j - C_{ijk} S^k + C_{ijk} S^{kl} K_l + h_{ij},
\end{eqnarray}
where 
\begin{equation} \label{gen}
  \der_{\bar{i}} S^{ij} = \bar{C}^{ij}_{\bar{i}}, ~ \der_{\bar{i}} S^j = G_{i \bar{i}} S^{ij}, ~ \der_{\bar{i}} S = G_{i \bar{i}} S^i, ~ K_i = \der_i K,
\end{equation}
and $h^{jk}_i$, $h^i_j$, $h_i$ and $h_{ij}$ denote holomorphic functions. The propagators $S^{ij}$, $S^i$ and $S$ are obtained as solutions
of the equations (\ref{gen}) up to holomorphic functions $f^i_{kl}$, $f_{kl}$ and $f$:
\begin{eqnarray} \label{propagators}
  S^{ij} & = & (C_k^{-1})^{jl}((\delta^i_k \der_l + \delta^i_l \der_k)K + \Gamma^i_{kl} + f^i_{kl}), \nonumber \\
  S^i    & = & (C_k^{-1})^{il}(\der_k K \der_l K - \der_k \der_l K + f^j_{kl} \der_j K) + f_{kl}), \nonumber \\
  S      & = & \frac{1}{2h^{11}}\left[(h^{1,1}+1) S^i - D_j S^{ij} - S^{ij} S^{kl} C_{jkl}\right]\der_i (K + \log(|f|)/2) \nonumber \\
  ~      & ~ & + \frac{1}{2h^{1,1}}(D_i S^i + S^i S^{jk} C_{ijk}),
\end{eqnarray}
where the matrix $C_k^{-1}$ is the inverse of the matrix $(C_k)_{ij} = C_{ijk}$.
The relations (\ref{trunc}) imply that the topological free energies $\cF^{(g)}$ are polynomials in a finite set of non-holomorphic generators, 
namely the propagators $S^{ij}$,~$S^i$,~$S$ and the K\"ahler derivatives $K_i$. To see this, note that equation (\ref{defF1}) can be written
in terms of these generators as
\begin{equation} \label{propF1}
  \der_i \cF^{(1)} = \frac{1}{2} C_{ijk} S^{jk} - (\frac{\chi}{24}-1)K_i + A_i,
\end{equation}
where the holomorphic ambiguity is encoded in the ansatz $A_i = \der_i(\tilde{a}_j \log \Delta_j + \tilde{b}_j \log z_j)$. Rewriting 
the left hand side of equation (\ref{holan}) as
\begin{equation}
  \bar{\der}_{\bar{\imath}} \cF^{(g)} = \bar{C}^{jk}_{\bar{i}} \frac{\der \cF^{(g)}}{\der S^{jk}} 
                              + G_{i\bar{\imath}} \left(\frac{\cF^{(g)}}{\der K_i} + S^i \frac{\der \cF^{(g)}}{\der S} 
                              + S^{ij} \frac{\der \cF^{(g)}}{\der S^j}\right),
\end{equation}
and assuming independence of the $\bar{C}^{jk}_{\bar{\imath}}$ and the $G_{i\bar \imath}$ gives
\begin{eqnarray} \label{Ftrunc}
  \frac{\der \cF^{(g)}}{\der S^{ij}} & = & \frac{1}{2} D_i D_j \cF^{(g-1)} + \frac{1}{2} \sum_{r=1}^{g-1} D_i \cF^{(g-r)} D_j \cF^{(r)}, \nonumber \\
  0                               & = & \frac{\der \cF^{(g)}}{\der K_i} + S^i \frac{\cF^{(g)}}{\der S} + S^{ij} \frac{\der \cF^{(g)}}{\der S^j}.
\end{eqnarray}

Due to (\ref{trunc}) and (\ref{propF1}) the right hand side of these equations is always a polynomial in the generators (\ref{gen}).
Therefore, it is straightforward to integrate the equations (\ref{Ftrunc}) which finally shows the polynomiality of the free energies.
The last equation in (\ref{Ftrunc}) can be used to show that  $\cF^{(g)}$ becomes independent of the $K_i$ in the redefined basis
\begin{eqnarray}
  \tilde S^{ij} & = & S^{ij}, \nonumber \\
  \tilde S^i   & = & S^i - S^{ij} K_j, \nonumber \\
  \tilde S     & = & S - S^i K_i + \frac{1}{2} S^{ij} K_i K_j, \nonumber \\
  \tilde K_i   & = & K_i,
\end{eqnarray}
and one has $\der \cF^{(g)}/ \der \tilde K_i = 0$.

For practical calculations it is convenient to work in the basis of the tilted generators and we therefore rewrite the 
truncation relations (\ref{trunc}) in terms of these as
\begin{eqnarray}
  D_i \tilde S^{kl} & = & \tilde S^l \delta^k_i + \tilde S^k \delta^l_i + \tilde K_j \tilde S^{jl} \delta^k_i + \tilde K_j \tilde S^{jk} \delta^l_i
                    - C_{imn} \tilde S^{km} \tilde S^{ln} + h^{kl}_i, \nonumber \\
  D_i \tilde S^k  & = & 2 \tilde S \delta^k_i + \tilde K_m \tilde S^m \delta^k_i 
                        - \delta^m_i \tilde K_m \tilde S^k - h_{im} \tilde S^{mk} + h^k_i, \nonumber \\
  D_i \tilde S    & = & -2 \tilde S \tilde K_i - h_{im} \tilde S^m + \frac{1}{2} C_{imn} \tilde S^m \tilde S^n + h_i , \nonumber \\
  D_i \tilde K_j  & = & - \tilde K_i \tilde K_j - C_{ijk} \tilde S^k + h_{ij}.
\end{eqnarray}

The holomorphic functions $h^{kl}_i$, $h^k_i$, $h_i$ and $h_{ij}$ are extracted from expansions of the above equations around the 
large complex structure point in moduli space and we refer the reader to section \ref{ModelSolutions} and 
equations (\ref{M1hamb}), (\ref{M2hamb}) for their concrete form.

\subsubsection{The holomorphic ambiguity}
\label{holamb}

As in the case of genus $1$ there also arise holomorphic ambiguities at higher genus due to the anti-holomorphic derivative in (\ref{holan}).
These ambiguities, denoted by $f_g$, are rational functions defined on the whole moduli space and transform as sections of $\cL^{2-2g}$.
One of the major challenges of topological string theory is to fix the ambiguity at each genus, after each integration step. This is done 
through using physical boundary conditions at the boundary divisors of the moduli space. In \cite{integrability} it was shown that 
the gap condition near conifold singularities found in \cite{HKQ} provides always enough information to fix all constants parametrizing
the ambiguities $f_g$ in the case of local Calabi-Yau manifolds. There, regularity at the orbifold point and at the large radius point,
as well as the behaviour near the conifold implied the following ansatz for $f_g$:
\begin{equation} \label{fglocal}
  f_g = \sum_i \frac{A^i_g}{\Delta_i^{2g-2}}.
\end{equation}
The $A_g^i$ are polynomials in $z$ of degree $(2g-2) \cdot \textrm{deg} \Delta_i$ and the sum runs over all irreducible 
components of the discriminant locus. However, in the case of compact Calabi-Yau manifolds one is dealing with several boundary divisors and many of them arise through a blow up of the moduli space and do not manifest themselves as singular loci of the Calabi-Yau hypersurface. A convenient way to see what is happening around these divisors and whether they have to be introduced in the holomorphic ambiguities for higher genera is to look at the behaviour of the genus $1$ free energy $F_i^1(t_{i,N},t_{i,T})$. Here, $i$ is a label for the divisor in question and $t_{i,N}$, $t_{i,T}$ denote the flat coordinates normal as well as tangential to the divisor. In the work of Vafa \cite{ConiFate} it is argued that the coefficient in front of the term logarithmic in $t_{i,N}$ counts the difference between hyper- and vectormultiplets which become massless at the divisor $\Delta_i$, i.e. we have the following expansion
\begin{equation}
  F^1_i(t_{i,N},t_{i,T}) = (n_H - n_V) \log(t_{i,N}) + \cdots .
\end{equation}

This allows us to constrain the form of the ambiguity for higher genera by demanding regularity at all divisors whose corresponding $F^1$-expansion does not come with a logarithmic term in the normal direction. This path of argumentation leads us to the following ansatz for the holomorphic ambiguities
\begin{equation} \label{fgcompact}
  f_g = \sum_{|I|\leq P_{\infty}(g)} a_I z^I + \sum_k \frac{\sum_{|I|\leq P_{k}(g) \cdot \textrm{deg} \Delta_k} c^k_I z^I}{\Delta_k^{P_k(g)}},
\end{equation}
where $z^{I}$ is a short hand notation for $z_1^{i_1} z_2^{i_2} \cdots z_n^{i_n}$ and $|I| = i_1 + \cdots i_n$.
Furthermore, $P_k(g)$ denotes the power of the boundary divisor $\Delta_k$ as a function of the genus $g$. Note that we also have terms which are polynomial in the $z_i$ and therefore become singular around the locus $\Delta_{\infty}$ where $z_i \rightarrow \infty$. 

The power of $\Delta_k$ in the
denominator is fixed by the leading behaviour of $F^g$ near the corresponding singularity. In the case of the conifold
singularity the behaviour is of the form
\begin{equation} \label{gap}
  F^g_c = \frac{c^{g-1} B_{2g}}{2g(2g-2)t_{c,N}^{2g-2}} + \cO(t_c^0),
\end{equation}
where $t_{c,N} \rightarrow 0$ is a flat coordinate normal to the singularity locus. For more general singularities where $n_H$ hypermultiplets and $n_V$ vectormultiplets become massless one expects the behaviour  
\begin{equation} \label{generalgap}
  F^g_s = (n_H-n_V) \frac{c^{g-1} B_{2g}}{2g(2g-2)t_{s,N}^{2g-2}} + \cO(t_s^0),
\end{equation}
where $t_{s,N}$ is again the coordinate normal to the singularity locus. In order to extract the power of the discriminant component in the denominator of the ansatz one has to take into account the relation between $\Delta_s$ and $t_{s,N}$. In the case of the conifold discriminant this behaviour is a direct proportionality which is the reason why this discriminant appears to inverse powers of $2g-2$. In the case of the strong coupling discriminant we will be dealing in our examples the relation is $\Delta_s \sim t_{s,N}^2$ which leads to the ansatz
\begin{equation}
  f_g = \ldots + \frac{\sum_{|I| \leq g-1} c^s_{I} z^I}{\Delta_s^{g-1}} + \ldots.
\end{equation}

Formula (\ref{generalgap}) can be traced back to the observation of Gopakumar and Vafa \cite{MtopII} that topological string amplitudes can be alternatively computed
by a supersymmetric version of the Schwinger loop calculation where
BPS particles are running in the loop. As a half-vectormultiplet contains
one more fermion than a half-hypermultiplet there is a relative minus sign between the two loop calculations. Another way to see this is that in $N=4$ theories where one has no quantum corrections $N=2$ vector- and hypermultiplets are forming together one $N=4$ multiplet. Therefore, 
one expects that quantum corrections come with a sign difference. 

However, the above argumentation leading to the result (\ref{generalgap}) goes only through once the theory is noninteracting and the calculation for the various BPS particles can be done separately. In an interacting theory several BPS states can form a bound state and then the calculation of the effective field theory becomes much more involved. In our case we find that the theory at the strong coupling divisor $\Delta_s$ is interacting as genus $4$ computations exclude a simple gap as in (\ref{generalgap}). Therefore, in order to deduce the correct boundary conditions an effective field theory calculation has to be performed. 

Note that in the case of local Calabi-Yau manifolds the vanishing of subleading terms in (\ref{gap})
provides enough boundary conditions in order to fix the $A^i_g$ and therefore the ambiguity completely \cite{integrability}. 
We claim that this is also the case in the compact examples, i.e. that the constants $c^k_I$ are fixed completely by the leading behaviour near the corresponding singularity $\Delta_k$. This fact was already observed in the case of conifold singularities in \cite{GrCY}. 
However, once we are dealing with compact manifolds also terms of the form $a_I z^I$ appear in the ambiguity which become singular near the divisors $z_i = \infty$. 
The constant term in this series is always solved for by the constant map contribution to $F^g$ at the point of large radius in moduli space
\begin{equation}
  F^g = \frac{\chi B_{2g-2} B_{2g}}{4g (2g-2)(2g-2)!} + \cO(e^{2\pi i t}).
\end{equation}
The terms linear and of higher order in the $z_i$ seem to be connected to new physics becoming important at the divisors $z_i=\infty$ and/or their intersections.
In the case of K3 fibrations relevant for this paper we find that there are two kinds of information provided at these boundary divisors. First of all, in our models,
at $z_2 = 0$ which is the point of large base volume heterotic/type II duality can be used to calculate fibre invariants which will impose constraints on the $a_I$.
Secondly at the intersection point $z_1 = z_2 = \infty$ one has an exact CFT description and thus 
one can impose regularity on the amplitudes around this point. We find that $P_{\infty}(g) = \lfloor\frac{3}{2}(g-1)\rfloor$ and that the two conditions mentioned 
provide enough information to solve for the constants $a_I$ up to genus $4$ and most probably to all genera.

\section{Moduli Space of K3 Fibrations}
\label{K3ModuliSpace}

In this section we want to give an overview of the moduli Space of K3 fibrations as presented in \cite{CandelasMirrorII}. Special attendance will be given to 
boundary divisors and their importance for physical boundary conditions. 

\subsection{K3 Fibrations}
\label{K3Fibrations}

K3 fibrations arise in the context of heterotic/type II duality once one wants to have $N=2$ supersymmetry in four dimensions \cite{KachruVafa}. 
In order to achieve this amount of supersymmetry on the heterotic side one has to compactify on $K3 \times T^2$. In the heterotic picture
vector multiplets come from the 2-torus together with its bundle and the dilaton axion, while hypermultiplets arise as deformations of the
K3 surface and its bundle. On the type II side vector multiplets arise from compactification of the R-R three-form and therefore count $h^{1,1}(M)$,
while the complex structure deformation parameters of the Calabi-Yau together with the dilaton-axion form $h^{2,1}(M)+1$ hypermultiplets. 
Part of the duality conjecture is that there is a complete match between the moduli spaces of the two theories. To see the consequences it is 
easiest to start with the vector multiplet moduli of the heterotic side. Here one has from the Narain moduli of the 2-torus and the dilaton-axion
locally a product of the form
\begin{equation} \label{HeteroticModuliSpace}
  \frac{O(2,m)}{O(2)\times O(m)} \times \frac{SL(2)}{U(1)}.
\end{equation}
The classical vector moduli space $\cM_V$ of the type IIA theory is a special K\"ahler manifold. Choosing special coordinates,
the K\"ahler potential takes the following form in terms of the prepotential $F$
\begin{eqnarray} \label{SpecialKaehler}
  K            & = & - \log\left( 2(F + \bar{F}) - (t^i - \bar{t}^i)\left(\frac{\der F}{\der t^i} - \frac{\der \bar{F}}{\der \bar{t}^i}\right)\right), \nonumber \\
  G_{i\bar{j}} & = & \frac{\der K}{\der t^i \der \bar{t}^j}. 
\end{eqnarray}
Here, $G_{i\bar{j}}$ is the metric on moduli space. Imposing the local product structure (\ref{HeteroticModuliSpace})
on a special K\"ahler manifold allows one to deduce the prepotential from equations (\ref{SpecialKaehler}). 
This, on the other hand, opens up the way for calculating intersection numbers 
$\kappa_{ijk} = \sharp(D_i \cap D_j \cap D_k)$ from the special K\"ahler relation
\begin{equation}
  \kappa_{ijk} = \frac{\der F}{\der t_i \der t_j \der t_k}.
\end{equation}
Such a calculation was performed in \cite{Ferrara} and the result is
\begin{eqnarray} \label{K3fibration}
  \sharp(D_0 \cap D_0 \cap D_0) & = & 0, \nonumber \\
  \sharp(D_0 \cap D_0 \cap D_i) & = & 0,~~ i=1, \cdots, h^{1,1} - 1 \nonumber \\
  \sharp(D_0 \cap D_i \cap D_j) & = & \eta_{ij},~~ i,j=1, \cdots, h^{1,1}-1, 
\end{eqnarray}
where $\eta_{ij}$ is a matrix of nonzero determinant and signature $(+,-,-,\cdots,-)$. A theorem of Oguiso \cite{Oguiso} 
says that if $M$ is a Calabi-Yau threefold and $L$ a divisor such that
\begin{equation} \label{K3cond}
  L \cdot c \geq 0 \textrm{ for all curves } c \in H_2(M,\Z), \quad L^2 \cdot D = 0 \textrm{ for all divisors } D \in H_4(M,\Z),
\end{equation}
then there is a fibration $\Phi: M \rightarrow W$, where $W$ is $\P^1$ and the generic fibre $L$ is either  a K3 surface or an abelian
surface. The second condition of (\ref{K3cond}) follows from (\ref{K3fibration}) with $L = D_0$. Further investigation \cite{AspLouis}
shows that the first condition in (\ref{K3cond}) is also true and that the Euler characteristic of the fibre given by the second
chern class is $24$. This determines the fibration as a K3 fibration over $\P^1$, where the first factor of (\ref{HeteroticModuliSpace}) arises
from the Picard-lattice of the K3 fibre and the second factor is identified with the K\"ahler form plus $B$-field on $\P^1$. Therefore we see
that in this picture the heterotic dilaton is identified with the size of the $\P^1$ on the type IIA side.

\subsection{The Moduli Space of the Mirror}

In this section we will concentrate on the example $M_1 = \P_4^{(1,1,2,2,2)}[8]$ whose mirror moduli space we will describe. The case of 
$M_2 = \P_4^{(1,1,2,2,6)}[12]$ is analogous. $M_1$ is a generic degree 8 hypersurface in the weighted projective space $\P_4^{(1,1,2,2,2)}$. A 
typical defining polynomial for such a hypersurface is 
\begin{equation}
  p = x_1^8 + x_2^8 + x_3^4 + x_4^4 + x_5^4.
\end{equation}
One sees that there is a $\Z_2$-singularity along $[0,0,x_3,x_4,x_5]$. This may be blown up, replacing each point in the locus by $\P^1$
with homogeneous coordinates $[x_1,x_2]$ which will be the base of the fibration.  Choosing a point on the base by fixing $x_1/x_2=\lambda$
projects onto the subspace $\P_3^{(1,2,2,2)}$ with the fibre given by the hypersurface
\begin{equation}
  (\lambda^8+1)x_2^8 + x_3^4 + x_4^4 + x_5^4 = 0.
\end{equation}
This is seen to be a quartic K3 surface in $\P_3$ ones one makes the substitution $y_1 = x_2^2$.
Thus we see that $M_1$ is a K3 fibration with two K\"ahler moduli $t_1$ and $t_2$, $t_2$ being the
size of the $\P^1$ base while $t_1$ corresponds to the curve cut out by a generic hyperplane in the 
K3 fibre.
Its mirror $W_1$ may be identified with the family of Calabi-Yau threefolds of the form $\{p=0\}/G$, where
\begin{equation} \label{W1eq}
  p = x_1^8 + x_2^8 + x_3^4 + x_4^4 + x_5^4 - 8 \psi x_1 x_2 x_3 x_4 x_5 - 2 \phi x_1^4 x_2^4.
\end{equation}
Here, $G$ consists of elements $g = (\alpha^{a_1}, \alpha^{a_2}, \alpha^{2a_3}, \alpha^{2 a_4}, \alpha^{2a_5})$ with
the action
\begin{equation}
  (x_1, x_2, x_3, x_4, x_5; \psi, \phi) \mapsto 
  (\alpha^{a_1} x_1, \alpha^{a_2} x_2, \alpha^{2a_3} x_3, \alpha^{2a_4} x_4, \alpha^{2a_5} x_5; \alpha^{-a} \psi, \alpha^{-4a} \phi),
\end{equation}
where $a = a_1 + a_2 + 2a_3 + 2 a_4 + 2 a_5$, where $\alpha^{a_1}$ and $\alpha^{a_2}$ are $8$th roots of unity, and where 
$\alpha^{2a_3}$, $\alpha^{2a_4}$, and $\alpha^{2a_5}$ are $4$th roots of unity. 
Therefore, we see that the parameter space $\{(\psi, \phi)\}$ is modded out by a $\Z_8$ acting in the form
\begin{equation}
  (\psi, \phi) \mapsto (\alpha \psi, -\phi).
\end{equation}
This translates to the description of the moduli space as an affine quadric in $\C^3$
\begin{equation}
  \tilde \xi \tilde \zeta = \tilde \eta^2,
\end{equation}
with invariant coordinates
\begin{equation}
  \tilde \xi = \psi^8, \quad \tilde \eta = \psi^4 \phi, \quad \tilde \xi = \phi^2.
\end{equation}
Compactification of this space leads to the projective quadric $Q= \{\xi \zeta - \eta^2=0\}$ in $\P_3$ with coordinates 
$[\xi, \eta, \zeta, \tau]$.
The relation to the tilted coordinates is $\tilde \xi = \xi/\tau,~ \tilde \eta =\eta/\tau,~ \tilde \zeta = \zeta/\tau$
for $\tau \neq 0$.
Having identified the global form of the moduli space let us now proceed to the description of boundary divisors which
correspond to parameter values for which the original family of hypersurfaces $\{p=0\}$ develops singularities.
The analysis was done in \cite{CandelasMirrorII} and the result is the following set of boundary divisors:
\begin{eqnarray}
  C_{con}    & = & Q \cap \{1 - 2 x + x^2 (1 - y) = 0\},\\
  C_1        & = & Q \cap \{1-y = 0\},\\
  C_{\infty} & = & Q \cap \{y = 0\} : ~~ \phi \textrm{ and } \psi \textrm{ both approach infinity},\\
  C_0        & = & Q \cap \{1/x = 0\}.
\end{eqnarray}

Here we have used the coordinates \footnote{For the model $\P_4^{(1,1,2,2,6)}[12]$ use $x := -\frac{1}{864} \frac{\phi}{\psi^6}$, $y = \frac{1}{\phi^2}$.}
\begin{equation} \label{xyvar}
  x := - \frac{1}{8} \phi \psi^{-4}, \quad y = \phi^{-2},
\end{equation}

which are themselves related to the coordinates $z_i$ in (\ref{zmoduli}) through $z_1 = x/256$ and $z_2 = y/4$.
This choice is convenient for the description of the mirror map which becomes
\begin{equation} \label{mirrormap}
  t_1 \sim \log(x), \quad t_2 \sim \log(y).
\end{equation}

\begin{figure}
  \begin{center}
    \includegraphics[scale=0.8]{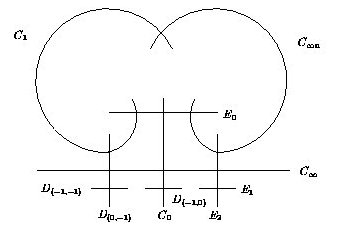}
    \caption{The blown up moduli space}
    \label{FigMS}
  \end{center}
\end{figure}

$C_{con}$ corresponds to the locus where the Calabi-Yau developers a conifold singularity and along $C_1$
the Calabi-Yau manifold $M_1$ admits a whole singular curve of genus $g$ over which $A_{n-1}$ singularities are fibred
(in our particular case $g = 3$ and $n=2$). $C_{\infty}$ is the locus where the volume of the $\P^1$ base
goes to infinity.  
Next, one notices that the resulting space is singular. First of all the quadric $Q \subset \P^3$ is singular
by itself. This singularity is of toric origin as $Q$ can be identified isomorphically with $\P^{1,1,2}$.
Further singularities arise from the point of tangency between the divisors $C_{con}$, $C_{\infty}$, from the tangency
between $C_1$, $C_{\infty}$ (of toric origin), and from the common point of intersection of $C_0$, $C_1$ and $C_{con}$.
Blowing up all singular points leads to the schematic picture of the moduli space presented in figure \ref{FigMS}.

\subsection{Physical boundary conditions}
\label{PhysBound}

\subsubsection{The Strong coupling singularity}
\label{StrongCoupling}

Consider the locus $C_1 = Q \cap \{\phi^2 = 1 \}$. It can be shown that the mirror map converts this locus to the locus $t_2 = 0$ in 
the K\"ahler moduli space of the Calabi-Yau $M_1$ \cite{MayrKlemm}. As $t_2$ describes the size of the $\P^1$ which is the base
of the K3 fibration $t_2 = 0$ translates to the strong coupling regime in the dual heterotic picture. In $M_1$ the singularity
is described by the equations
\begin{equation}
  x_1 = x_2 = 0 , \quad x_3^4 + x_4^4 + x_5^4 = 0,
\end{equation}
leading to a genus $3$  curve \footnote{In $M_2$ the equations describing the singularity lead to a genus 2 curve}
 $C$ of fixed points of the projective action $x_i \mapsto \mu^{Q_i} x_i$. In the language of toric
geometry the singular curve $C$ corresponds to a one-dimensional edge of the dual polyhedron $\Delta^*$ with integral lattice points
on it. The resolution process adds a new vertex for each of these points leading to an exceptional $\P^1$ bundle over $C$ in the
blown up of the Calabi-Yau manifold for each ray added. The monomial divisor mirror map relates each vertex to the addition
of a new perturbation in the defining polynomial of $W_1$. In our case we blow up only once and the perturbation added 
is the term $\phi x_1^4 x_2^4$ in (\ref{W1eq}).

To see what happens from the physics point of view along $C_1$ we look at the effective action arising from compactification in the type IIA picture.
This procedure has been analyzed in \cite{EnhancedGS}. Let us first clarify the setup. Assume that we have a smooth curve 
of genus $g$ and singularities of type $A_{N-1}$ fibred over the curve. The resolution of the transverse $A_{N-1}$ singularity
gives rise to an ALE space in which the vanishing cycles are described by a chain of $N-1$ two-spheres $\Gamma_i$ and 
their intersection matrix corresponds to the Dynkin diagram of $A_{N-1}$. Now consider soliton states described by two-branes 
wrapping the two-cycles $A^{ij}$ defined by the chain $\Gamma_i \cup \Gamma_{i+1} \cup \cdots \cup \Gamma_j$. These become charged
under $U(1)^{N-1}$ with their charges being identified with the positive roots of $A_{N-1}$. Compactification of the theory down
to $4$ dimensions leads to a $N=2$ supersymmetric $SU(N)$ gauge theory with $g$ hypermultiplets (coming from holomorphic 
1-forms on $C$) transforming in the adjoint
representation of the gauge group. In $N=1$ superfield notation, we obtain the following effective Lagrangian
\begin{equation}
  2 \pi \cL = \textrm{Im} \left[\textrm{Tr} \int d^4 \theta ({M_i}^{\dagger} e^V M^i + \tilde {M^{\dagger}}^i e^V \tilde M_i 
              + \Phi^{\dagger} e^V \Phi) + \frac{\tau}{2} \int d^2 \theta \textrm{Tr} W^2 + i \int d^2 \theta \mathcal{W}\right],
\end{equation}
with the superpotential
\begin{equation}
  \cW = \textrm{Tr} \tilde M^i [\Phi, M_i],
\end{equation}
and the scalar potential
\begin{eqnarray} \label{SCaction}
  \mathcal{V} & = & \textrm{Tr} \left[[m_i, {m^{\dagger}}^i]^2 + [\tilde m^i, {\tilde m_i}^{\dagger}]^2 + [\phi,\phi^{\dagger}]^2\right. \nonumber \\
     ~        & ~ &  + \left. 2 \left([{m^{\dagger}}^i,\phi] [\phi^{\dagger},m_i] + [{\tilde m^{\dagger}}_i,\phi][\phi^{\dagger},\tilde m^i] 
                     + [m_i, \tilde m^i][{\tilde{m}^{\dagger}}_j, {m^{\dagger}}^j]\right)\right].
\end{eqnarray}

Here $V_a$ is the vector multiplet in the adjoint and $W^a_{\alpha}$ its field strength. Furthermore, one has a chiral superfield $\Phi^a$
in the adjoint (comprising with $V$ the $N=2$ vector multiplet), and $2g$ chiral superfields $M^i_a$, $\tilde M^i_a$
in the adjoint (comprising the $g$ hypermultiplets, $i=1,\cdots,g$). Going to the Coulomb branch of the moduli space, 
$\phi = \textrm{diag}(\phi_1, \phi_2, \cdots, \phi_N)$ with $\sum \phi_i = 0$, one sees that at generic points along this 
the gauge symmetry is spontaneously broken to $U(1)^{N-1}$. In codimension one (along the singular divisor $C_1 \sim \phi_N=0$) the unbroken 
symmetry gets enhanced to $SU(2) \times U(1)^{N-2}$ and from (\ref{SCaction}) one can deduce that $2g$ hypermultiplets and $2$ vectormultiplets
are becoming massless near the locus $\phi_N = 0$. Therefore, the number $n_H - n_V$ in (\ref{generalgap}) becomes $2g-2$.

\subsubsection{The weak coupling divisor and meromorphic modular forms}
\label{WeakCoupling}

The weak coupling divisor deserves its name from the definition $y = 0$ which inserted into the mirror map (\ref{mirrormap}) gives 
$t_2 \rightarrow \infty$. Again as $t_2$ describes the size of the dilaton in the heterotic dual we are in the weak coupling regime
of the heterotic string. This can be used to calculate higher genus amplitudes in the type IIA string through a heterotic one loop
computation. As was shown in \cite{NarainI} topological string theory calculates certain F-terms in the four dimensional effective
field theory. These are graviton - graviphoton couplings of the form
\begin{equation} \label{Fterms}
  \int d^4 x d^4 \theta \mathcal{W}^{2g} F^g(X) = \int d^4 x F^g(t_i) R_+^2 F_+^{2g-2} + \cdots,
\end{equation}
where $R_+$ is the self-dual part of the Riemann tensor, $F_+$ the self-dual part of the graviphoton field strength and $F^g$ denotes the
topological free energy at genus $g$. As the $F^g(X)$ are homogeneous functions of $X^I$'s of degree $2-2g$
and $X^0$ can be chosen to be
\begin{equation}
  X^0 = \frac{1}{g_s} e^{K/2},
\end{equation}
one can write
\begin{equation} \label{Fgdilaton}
  F^g(X) = (X^0)^{2-2g} F^g(t) = (g_s^2)^{g-1} e^{(1-g)K} F^g(t).
\end{equation}
In type IIA string theory the K\"ahler potential $K$ is independent of the dilaton since the latter belongs to a hypermultiplet and
there are no neutral couplings between vector multiplets and hypermultiplets. The same argument tells us that $F^g(t)$ is independent
of the dilaton and it follows from (\ref{Fgdilaton}) that the couplings (\ref{Fterms}) appear only at genus $g$. Switching to the 
heterotic picture this statement changes as follows. Now the K\"ahler potential contains a $\log(g_s^2)$ term. This term arises from
the vector moduli prepotential of the type IIA theory 
\begin{equation}
  F \sim ST^2 + \sum_{n=0}^{\infty} f_n(T) \textrm{exp}(-nS)
\end{equation}
with the identifications $T = t_1$, $S = t_2$ and the choice $S = \frac{\theta}{2\pi} + i \frac{8 \pi}{g_s^2}$. 
Next, notice that this implies that $X^0$ is of order $1$ in the dilaton and therefore one extracts from (\ref{Fgdilaton}) 
that all $F^g$ appear at one loop in the heterotic theory. However, in the case of the heterotic string the dilaton
belongs to a vector multiplet and therefore all $F^g(t)$ can have nontrivial dilaton dependence apart from the dependence
through $X^0$. This implies that the 
above analysis is only valid in the limit $S \rightarrow \infty$ where the dilaton dependence in the $F^g(t)$ drops out.
Translated to the type II picture we see that the one loop calculation gives only control over the terms in $F^g(t)$ 
which are independent of the class of the $\P^1$ base $t_2$. 
Such a calculation was performed in \cite{NarainII,MarinoMoore} and extended
to arbitrary regular K3 fibrations in \cite{KlemmScheidegger}. The result is
\cite{KlemmScheidegger} that the Gopakumar-Vafa invariants for the K3 fibre are encoded in the following generating function
\begin{equation}
  F_{K3}(\lambda,q) = \frac{2 \Phi_{N,n}(q)}{q} \left(\frac{1}{2 \sin(\frac{\lambda}{2})}\right)^2 \prod_{n\geq 1}
                      \frac{1}{(1-e^{i\lambda} {q}^n)^2 (1-{q}^n)^{20} (1 - e^{-i\lambda} {q}^n)^2}, 
\label{fibre}
\end{equation}
where $q = e^{2\pi i \tau}$ and $\Phi_{N,n}(q)$ is a modular form of half integral
weight with respect to a congruent subgroup of $PSL(2,\mathbb{Z})$ acting in the standard 
form on $\tau$. Let us review the essential points of this formula 
before we specify $\Phi_{N,n}(q)$ explicitly for the relevant class of 
K3 fibrations.    

The formula applies to multi parameter K3 fibrations such as the 3 parameter  
STU model, as the Gopakumar-Vafa invariant depends on the class $[C]$ of the 
curve only via the self intersection $C^2$, and the latter is related to 
the exponents of the parameter $q$. 

$\Phi_{N,n}$ is fully determined by the genus zero Gopakumar-Vafa invariants 
of the fibre direction. As it has been pointed out in
\cite{MaulikPandharipande}, 
it can be also determined by classical geometric 
properties of the fibration, namely  the embedding  
\begin{equation}
  \iota : \textrm{Pic}(K3) \hookrightarrow H^2(M,\Z),
\label{embeddings}
\end{equation}
and the Noether-Lefshetz numbers of regular, i.e. non singular, 
one parameter families of quasi-polarized K3 surfaces 
$\pi:X \rightarrow {\cal C}$, where $\cC$ is a curve. Let $L$ be a
quasi-polarization of degree  
\begin{equation} 
\int_{K3} L^2=2 N.
\end{equation} 
Then the family $\pi$ yields a morphism $\imath:{\cal C}\rightarrow 
{\cal M}_{2N}$ to the moduli space of quasi-polarized K3 
surfaces of degree $2N$. The Noether-Lefshetz numbers 
are defined by the intersection of ${\cal C}$ with the 
Noether-Lefshetz divisors in ${\cal M}_{2N}$. The 
latter are the closure of the loci in  ${\cal M}_{2N}$ 
where the the rank of the Picard lattice is two. If 
$\beta$ is an additional class in $\textrm{Pic}(K3)$ the 
Noether-Lefshetz divisor $D_{h,d}\in {\cal M}_{2N}$ may 
be  labeled by $\int_{K3} \beta^2=2 h-2$ and 
$\int_{K3} \beta\cdot L=d$ and combined into a generating function. 
The seminal work of Borcherds~\cite{Bocherds} relates these generating functions 
of the Noether-Lefshetz numbers to modular forms using the 
relations between Heegner- and Noether-Lefshetz divisors.  
In fact one knows  that they are combinations of meromorphic 
vector valued modular forms of half integral weight. 
The theory of Borcherds can be viewed as a further extension of the work 
of Hirzebruch and Zagier on the modularity of counting functions 
of divisors in Hilbert modular surfaces.

In~\cite{KlemmScheidegger} a formula for $\Phi_{N,n}(q)$ of weight 
$\frac{21}{2}$ was found for regular K3 fibration Calabi-Yau, if 
the fibre is a quartic in $\mathbb{P}^3$, i.e. $N=2$, and if the fibre is a sixtic in 
the weighted projective space $\mathbb{P}(1,1,1,3)$, i.e. $N=1$.
In fact there is a general formula, which encorporates not only 
the $N$ parameter, but also a second parameter $n$ parametrizing 
the different embeddings (\ref{embeddings}). For the $N=1$ examples we have, \cite{Kawai},
\begin{equation}
\Phi_{1,n}(q)=
U E_4\biggl(\frac{U^4\,\left(39\,V^8 +26\,U^4\,V^4  -U^8\right) }{2^5}+
n\frac{V^4\,\left( 7\,U^8 - 6\,U^4\,V^4 - V^8 \right) }{2^7}\biggr)
\label{phi1}
\end{equation}
with $U=\theta_3(\frac{\tau}{2})$, $V=\theta_4(\frac{\tau}{2})$ in terms of 
Jacobian theta functions 
\begin{equation}
\theta_2 = \ds{\sum_{n=-\infty}^\infty q^{\frac{1}{2}(n + 1/2)^2}},\qquad 
\theta_3 = \ds{\sum_{n=-\infty}^\infty q^{\frac{1}{2}n^2}}, \qquad 
\theta_4 = \ds{\sum_{n=-\infty}^\infty (-1)^n q^{\frac{1}{2} n^2}}
\end{equation}
and $E_4=1-240 q+\ldots $ is the weight $4$ Eisenstein series.
According to (\ref{fibre}) this has the following $q$ expansion for the genus zero 
invariants 
\begin{equation}
\begin{array}{rl}
{2 \Phi_{1,n} \over \eta^{24}}(q^4)=
&\frac{2}{q^4} \!-\! 252 \!-\! 2496q \!-\! 223752q^4 \!-\! 725504q^5\! -\! 15530000q^8 \!-\! 38637504q^9 \ldots\\
&+ n\left( q^{-3} \!-\! 56 \!+\! 384 q \! - \! 15024 q^4 \!+\! 39933 q^5 \!-\! 523584 q^8 \!+\! 1129856 q^9\ldots  \right)\ 
\end{array}
\label{f1}
\end{equation}
and the coefficients of the $q^{d^2/N}$ are the genus zero BPS numbers 
$n^0_d$. We note in particular that the constant term is known from 
physical arguments and enumerative geometry to be the Euler number of 
the Calabi-Yau, i.e.
\begin{equation}
\chi=-252- n \,56\ .
\end{equation}  
The fibrations discussed in this paper belong to the $n=0$ case, but 
several manifolds with values $n\in \mathbb{Z}$ are realized 
as complete intersections or hypersurfaces in toric ambient 
spaces.

For the second type of fibrations that we treat in this paper with 
$N=2$ one has 
\begin{equation}
\begin{array}{rl}
\Phi_{2,n}=&
\! \! \! \frac{1}{2^{21}}(81\, U^{19} V^2 -3\, U^{21} + 627\, U^{18} V^3 + 14436\, U^{17} V^4 + 20007\,  U^{16} V^5 + 169092\, U^{15} V^6\\ & 
+ 120636\, U^{14} V^7 + 621558\,  U^{13} V^8 + 292796\, U^{12} V^9 + 1038366\,U^{11} V^{10} \\ & 
+ 346122\, U^{10}\,  V^{11} + 878388\, U^9\,  V^{12} + 207186\,U^8\,  V^{13}+ 361908\,U^7\, V^{14} \\ & 
+ 56364\, U^6 V^{15} + 60021\,U^5 V^{16} +  4812\,U^4 V^{17} + 1881\,U^3 V^{18} + 27\,U^2 V^{19} - V^{21})\\ &-
\frac{n}{2^{22}}
U\,V\,{( U^2 - V^2 ) }^4 ( U^{11} - 21\,U^{10}\,V - 43\,U^9V^2 - 297\,U^8V^3
-158\,U^7V^4 - 618\,U^6V^5 \\ & 
-206\,U^5V^6 - 474\,U^4V^7 - 99\,U^3V^8 - 129\,U^2V^9 - 7\,U\,V^{10} + 3\,V^{11} ) 
\end{array}
\label{phi2}
\end{equation}
with $U=\theta_3(\frac{\tau}{4})$ and $V=\theta_4(\frac{\tau}{4})$ so that
\begin{equation}
\begin{array}{rl}
{2 \Phi_{2,n} \over \eta^{24}}(q^8)=
&\frac{2}{q^8} \!-\! 168\! -\! 640q\!-\! 10032q^4 \!-\! 158436q^8 \!-\! 288384q^9\! -\! 1521216q^{12}\! -\! 10979984q^{16} + \ldots\\
&+ n\left( \frac{2}{q^4} \! -\! 28\! +\! 64q\! - \! 328q^4 \! -\! 1808q^8\! +\! 2624q^9 \! - \!7656q^{12}\! -\! 27928q^{16} \ldots\right) \ .
\end{array}
\label{f2}
\end{equation}
Again the genus zero invariants can be read of from the $q^\frac{d^2}{N}$ term and the $d=0$ 
coefficient is related to the Euler number
\begin{equation}
\chi=-168- n  28\ .
\end{equation}  

One difficulty with the above approach is that due to $\eta^{24}$ in the denominator 
one has to make an ansatz for the $\Gamma(4N)$ modular forms in 
the numerator of very high weight, as it is apparent in formula (\ref{f2}), while the quotient 
is always in ${\cal M}^!_{-\frac{3}{2}}(\Gamma_0(4N))$ and can be represented 
within a much smaller ring, as was pointed out by Zagier.  The shriek stands for meromorphic, 
i.e. these forms are allowed to have arbitrary pole order at the cusp at $\tau \rightarrow i\infty$. 

Such half integral, meromorphic forms are denoted by ${\cal M}^!_{k+\frac{1}{2}}(\Gamma_0(4N))$ 
with $k\in \mathbb{Z}$. They can be labeled by their weight and the 
pole order at the cusp. For a given weight there are forms of arbitrary pole
orders, but our knowledge of the maximal pole order keeps the problem finite.
For $N=1$ one can start with $\theta=f^{1,0}_\frac{1}{2}=\theta_3(2 \tau)$\footnote{ 
Let us introduce the notation $f_{r}^{N,p}$, where we denote with $r$ the weight, with $p$ 
the pole order and $N$ labels the congruence subgroup as above. We reserve the
character $f$ for forms, which are of the form
$f_{r}^{N,p}=\frac{1}{q^p}+reg.$, while forms denoted by other characters can
have subleading poles.}  and for $N>1$
one has to use the right combination of vector valued half integral forms
which transform under the metaplectic representation of $\Gamma_0(4 N)$. Let 
\begin{equation} 
\phi_k=\sum_{m=-\infty}^\infty q^{2 N m+k}, \qquad k=0,\ldots, N 
\label{thetak} 
\end{equation}   
then we define 
\begin{equation}
\theta=\phi_0+\phi_N+\frac{1}{2} \sum_{k=1}^{N-1} \phi_k\ . 
\label{theta}
\end{equation} 
 
Different weights and pole orders can be constructed systematically by the
following operations~\cite{Zagier1}: 
\begin{itemize}
\item Multiplying $f_r$ with $j(4 N\tau)$, where $j$ is the total modular invariant of 
$\Gamma_0=PSL(2,\mathbb{Z})$, keeps the weight and shifts the pole order of the cusp at
 $i \infty$ by ${-4 N}$. 
\item  Taking derivatives $f_{r+2}=f_r E_2 (4 N\tau )- \frac{3}{N r} f_r'(\tau)$ will 
shift the weight by $2$ but keeps the pole order. 
\item Take the first Rankin-Cohen bracket $f_{k+l+2}=[f_k,E_l(4 N)]=
k f_k E'_l(4 N)-l E_l(4 N) f'_k$, where $E_4,E_6,E_8=E_4^2,$ etc. are holomorphic 
modular forms of $\Gamma_1$, shifts the weight  by $l+2$
but keeps the pole order. The bracket is build so that
it cancels the inhomogeneous transformation of the derivatives 
\begin{equation} 
D=\frac{1}{2\pi i} \frac{\dd }{\dd \tau}\ .
\end{equation} 
We denote  $f':=Df$ etc. Similarly the nth rank Rankin-Cohen bracket of 
modular forms $f,g$ of weight $k,l$ is 
\begin{equation}
[f,g]_n=\sum_{r=0}^n (-1)^r \left(k+n-1\atop n-r\right)
\left(l+n-1\atop r\right) D^r f D^{n-r} g \ .
\label{rankincohen}
\end{equation}
and is modular of weight $k+l+2 n$. Choosing $f=f_r$ and $g$ a holomorphic 
modular form changes the weight, but keeps the pole order. 
\item Dividing by $\Delta(4 N\tau)= q^{4N} \prod_{m=1}(1-(q^{4N})^m)^{24}$ lowers the 
pole order by $-(4 N)$ and the weight by $-12$.
\end{itemize}

It is clear that $\frac{2 \Phi_{1,n}}{\eta^{24}}(q^4)$ must be of the form 
$f_{-\frac{3}{2}}^{1,4}(q^4)$, which can be build as follows. First we 
construct $u^{1,0}_{\frac{5}{2}}=\theta E_2(4 \tau)-
6 \theta'(\tau)=1-10 q-70q^4-48 q^5+\ldots$. Form this we can get two 
elements $u^{1,4}_{-\frac{3}{2}}=\frac{\theta E_{10}(4 \tau)}{\Delta(4 \tau)}$ 
and $v^{1,4}_{-\frac{3}{2}}=\frac{ u^{1,0}_{\frac{5}{2}} 
E_{8}(4 \tau)}{\Delta(4 \tau)}$, which we combine into a combination 
$f_{-\frac{3}{2}}^{1,4}=(5 u^{1,4}_{-\frac{1}{2}}+v^{1,4}_{-\frac{1}{2}})/6$
for which the third order pole $\frac{1}{q^3}$ vanishes. Further we 
consider $f_{-\frac{3}{2}}^{1,3}=-\frac{1}{4}\frac{[\theta,E_8(4 \tau)]}{\Delta(4 \tau)}$. 
Using the results of Borcherds and matching a finite number of genus zero
invariants it can hence be proven that   
 \begin{equation} 
{2 \Phi_{1,n} \over \eta^{24}}(q^4)=2 f_{-\frac{3}{2}}^{1,4} 
+n f_{-\frac{3}{2}}^{1,3}\ .
\label{ff1}
\end{equation}

For the fibration with $N=2$ we use the general form (\ref{theta}). Then as 
before we construct $u^{2,0}_{\frac{5}{2}}=\theta E_2(8\tau)-3
\theta'(8\tau)$. In the next step we construct $4$ functions of weight
$-\frac{3}{2}$ with leading pole $\frac{1}{q^8}$:
$u^{2,8}_{-\frac{3}{2}}=\frac{\theta E_{10}}{\Delta}$,
$v^{2,8}_{-\frac{3}{2}}=\frac{u^{2,0}_{\frac{5}{2}}  E_8}{\Delta}$,
$w^{2,7}_{-\frac{3}{2}}=\frac{\theta E_8'-16 \theta' E_8}{\Delta}$ and
$x^{2,7}_{-\frac{3}{2}}=\frac{u^{2,0}_{\frac{5}{2}} E_6'-16 \theta'
E_6}{\Delta}$. One needs these in  order to subtract all subleading 
poles and to define $f_{-\frac{3}{2}}^{2,4}= \frac{1}{576}
w^{2,8}_{-\frac{3}{2}}+\frac{5}{864}x^{2,8}_{-\frac{3}{2}}$, 
$f_{-\frac{3}{2}}^{2,7}= (u^{2,8}_{-\frac{3}{2}}-v^{2,8}_{-\frac{3}{2}})/3-
8 f_{-\frac{3}{2}}^{2,4}$ and finally 
$f_{-\frac{3}{2}}^{2,8}= u^{2,8}_{-\frac{3}{2}}-f_{-\frac{3}{2}}^{2,7}-
2 f_{-\frac{3}{2}}^{2,4}$. Now we find the result for       
\begin{equation} 
{2 \Phi_{2,n} \over \eta^{24}}(q^8)=2 f_{-\frac{3}{2}}^{2,8} 
+2 n f_{-\frac{3}{2}}^{2,4}\ .
\label{ff2}
\end{equation} 
It is conceivable that there exists a more general family of regular 
$K3$ fibrations with the $N=2$ fibre type, which involve $m
f_{-\frac{3}{2}}^{2,7}$, but they have not been determined, yet (maybe one should
check in Kreuzers list for the Eulernumbers). 

The advantage of the method is that it gives the answer for all 
regular one parameter families of $K3$ in a systematic manner. 
Since the construction should be clear by now we list here only
the significant terms of the relevant $f^{3,n}_{-\frac{3}{2}}$ 
for the case with $N=3$.
\begin{equation} 
\begin{array}{rl}
f_{-\frac{3}{2}}^{3,3}&=q^{-3} - 2 + 6\,q - 12\,q^4 + 14\,q^9 +\\
f_{-\frac{3}{2}}^{3,8}&=q^{-8} - 27 + 56\,q + 214\,q^4 - 1512\,q^9 \\ 
f_{-\frac{3}{2}}^{3,11}&=q^{-11} - 54 - 134\,q + 924\,q^4 + 10098\,q^9\\
f_{-\frac{3}{2}}^{3,12}&=q^{-12} - 74 - 336\,q - 2730\,q^4 - 17680\,q^9\\
\end{array}
\end{equation}
We find that
\begin{equation}
\frac{2\Phi_{1,n}}{\eta^{24}}(q^{12})=2 f_{-\frac{3}{2}}^{3,12}-4 n f_{-\frac{3}{2}}^{3,3}
\label{almost}
\end{equation}
reproduces the BPS numbers of the $N=3$ fibrations. Examples with $n=0$  
are the complete intersection CY $X_{3,2}(1,1,1,1,1)$ over 
$\mathbb{P}^1$ as well as $X_{6,4}(1,1,2,2,2,2)$.

Let us finish the section with the description how to obtain from (\ref{fibre}) 
the actual topological string partition functions of the topological string 
in the fibre direction. If $\frac{d^2}{2N}=l\in \mathbb{Z}$ we replace 
the power $\lambda^{2g-2} q^l\rightarrow \frac{\lambda^{2g-2}}{(2 \pi i)^{3-2g}} Li_{3-2g}(e^{2 \pi i d t})$, where
$t$ is the K\"ahlerparameter of the fibre. All other $q^r$ powers are
dropped. With this information and the multicovering formula we get the
following higher genus BPS invariants in the fibre direction.

\begin{table}[h!]
\centering
\begin{tabular}[h]{|c|ccccccc|}
\hline
&$d=0$   &  1 &  2 &   3 &    4 &    5 &       6   \\
\hline
$g$ &       &       &    &    &     &      &               \\
0&252& 2496& 223752& 38637504& 9100224984& 2557481027520& 805628041231176\\ 
1&4&   0& -492& -1465984& -1042943520& -595277880960& -316194812546140\\ 
2&0& 0& -6& 7488& 50181180& 72485905344& 70378651228338\\ 
3&0& 0& 0& 0& -902328& -5359699200& -10869145571844\\ 
4&0& 0& 0& 0& 1164& 228623232& 1208179411278\\ 
5&0& 0& 0& 0& 12& -4527744& -94913775180\\ 
6&0& 0& 0& 0& 0& 17472& 4964693862\\ 
7&0&0& 0& 0& 0& 0& -152682820\\ 
8&0& 0& 0& 0& 0& 0& 2051118\\ 
9&0& 0& 0& 0& 0& 0& -2124\\ 
10&0& 0& 0& 0& 0& 0& -22\\
\hline
\end{tabular}
\caption{BPS numbers for the $N=1$ fibre  }
\end{table}

\begin{table}[h!]
\centering
\begin{tabular}[h]{|c|cccccccc|}
\hline
& $d=0$     &  1 &  2 &   3 &    4 &    5 &       6 &      7  \\
\hline
$g$ &       &       &    &    &     &      &      &            \\
0& 168& 640& 10032& 288384& 10979984& 495269504& 24945542832& 1357991852672\\ 
1&4& 0& 0& -1280& -317864& -36571904& -3478901152& -306675842560\\ 
2&0& 0& 0& 0& 472& 875392& 220466160& 36004989440\\ 
3&0& 0& 0& 0& 8& -2560& -6385824& -2538455296\\ 
4&0& 0& 0& 0& 0& 0& 50160& 101090432\\ 
5&0& 0& 0& 0& 0& 0& 0& -1775104\\ 
6&0& 0& 0& 0& 0& 0& 0& 4480\\ 
\hline
\end{tabular}
\caption{BPS numbers for the $N=2$ fibre  }
\end{table}

\subsubsection{The Seiberg-Witten plane}
\label{SeibergWitten}

The emergence of the Seiberg-Witten plane as a divisor in the type II moduli space can be seen best in the case of the Calabi-Yau
$\P_4^{(1,1,2,2,6)}[12]$.
In terms of the $T$ and $S$ moduli of the heterotic string one finds that the mirror map (\ref{mirrormap}) can be 
written as \cite{KachruVafa} \cite{CandelasMirrorII} 
\begin{equation} \label{dualitymap}
  x = \frac{1728}{j(T)} + \cdots, \quad y = \textrm{exp}(-S) + \cdots.
\end{equation}
This immediately translates to a powerful relation between the $SU(2)$ enhanced symmetry point of the heterotic model at $T=i$ 
and the singular point of the conifold locus on the type IIA side. Observe that $j(i) = 1728$ which inserted into the duality map (\ref{dualitymap})
gives $x = 1$. On the type II side this is the point of tangency between the conifold divisor $C_{con}$ and the weak coupling divisor $C_{\infty}$.
Blowing up this singularity twice through inserting two $\P^1$'s gives the picture in figure \ref{FigMS}. The divisor $E_2$ describes the physics
of the Seiberg-Witten plane \cite{SeibergWitten} for rigid $SU(2)$ Yang-Mills theory once we decouple gravity. In order to see this we will follow the analysis in
\cite{KlemmVafaHeterotic}. As in the 
Seiberg-Witten theory the variable $u = \textrm{tr}\phi^2$ vanishes at the $SU(2)$ point , it should be identified with $x-1$ to leading order 
\begin{equation}
  x = 1 + \alpha' u + \cO(\alpha'^2),
\end{equation}
where the powers of $\alpha'$ are chosen such as to make the above expansion dimensionally correct. The second coordinate $y$ is related to the 
$SU(2)$ scale $\Lambda$ and coupling constant $e^{-\hat S}$ through
\begin{equation}
  y = \alpha'^2 \Lambda^4 \textrm{exp}(-\hat S) =: \epsilon^2.
\end{equation}
The above identifications translate the conifold locus $(1-x)^2 -x^2 y = 0$ to 
\begin{equation}
  u^2 = \Lambda^4 \textrm{exp}(- \hat S).
\end{equation}
Decoupling gravity now means sending $\alpha' \rightarrow 0$. That is we construct the variables $x_1 = x^2 y/ (x-1)^2$
and $x_2 = (x-1)$ which to leading order in $\alpha'$ correspond to $1/{\tilde u}^2$ and $\epsilon \tilde u$ 
\footnote{Here, $\tilde u = u/(\Lambda^2 e^{-\hat{S}/2})$ is the correct dimensionless variable to use, see \cite{KlemmVafaHeterotic}}.
These variables describe the Seiberg-Witten plane consistently at the semiclassical limit $\tilde u = \infty$, at the massless monopole
points $ \tilde u = \pm 1$ and at the $\Z_2$ orbifold point $\tilde u = 0$. It was shown in \cite{KlemmVafaHeterotic}
that one obtains the rigid periods $a$, $a_D$ as a subset of the periods of the Calabi-Yau by specialization of the Picard-Fuchs system
to the semi-classical regime, $\tilde u \rightarrow \infty$:
\begin{equation}
  (1,S, \sqrt{\alpha'} a, \sqrt{\alpha'} a_D, \alpha' u, \alpha' u S).
\end{equation}

As at the monopole point a charged dyon gets massless we expect that in the limit where gravity becomes important this picture translates to
a charged black hole becoming massless. Therefore, we expect that the topological amplitudes at this point will admit the conifold expansion
(\ref{gap}).

\subsubsection{The Gepner point}
\label{Gepner point}

As Gepner found out \cite{Gepner}, there is a Calabi-Yau minimal-model correspondence at the Fermat point in moduli space of the mirror. This is the point where $\phi=\psi=0$ and
corresponds via the mirror map (\ref{mirrormap}) to the deep interior point in the moduli space of the Calabi-Yau $M$. In the picture \ref{FigMS}
this is the point of intersection of the divisors $C_0$ and $D_{-1,0}$. The CFT description arising here is
the tensor product of minimal models each at level $P_i$ such that 
\begin{equation}
  \sum_{j=1}^{d+2} \frac{3 P_j}{P_j+2} = 3d,
\end{equation}
where $d$ is the complex dimension of the Calabi-Yau. This implies that $D$, the least common multiple of the $P_j +2$, satisfies
\begin{equation}
  D = \sum_{j=1}^{d+2} \frac{D}{P_j+2}.
\end{equation}
Therefore, one can interpret the Calabi-Yau equation
\begin{equation}
  \sum_{j=1}^{d+2} x_j^{P_j+2} = 0
\end{equation}
in the weighted projective space $\P^{d+1}(\frac{D}{P_1+2},\cdots, \frac{D}{P_{d+2}+2})$ as the superpotential of a Landau-Ginzburg theory
with chiral superfields $x_i$ \cite{GVW}(see \cite{WittenPhases} for a more rigorous description) . Then the conformal field theory description 
arises as the infrared fixed point of this theory.

The impact on the topological free energies $F^g$ is that these have to be regular in an expansion around the CFT point imposing boundary
conditions on the holomorphic ambiguity and in particular on the constants $a_I$ appearing in (\ref{fgcompact}).

\section{Solution of the Models}
\label{ModelSolutions}

In this section we present the results of our calculations for the models $\P_4^{(1,1,2,2,2)}[8]$ and $\P_4^{(1,1,2,2,6)}[12]$.

\subsection{$M_1 = \P_4^{(1,1,2,2,2)}[8]$}
\label{M1}

The toric data describing the ambient space of this Calabi-Yau can be summarized in the following table, where by $V$ we denote the collection of vectors 
$\bar{v}=(v,1)$ with $v$ being the integral points of the polyhedron $\Delta^*$ and $L$ are the corresponding charge vectors.

\begin{equation}  \label{vectorsM1}
  (V|L) = \left(\begin{array}{ccccc|cc}
                  -1 & -2 & -2 & -2 & 1 &  0 &  1 \\
                   1 &  0 &  0 &  0 & 1 &  0 &  1 \\
                   0 &  1 &  0 &  0 & 1 &  1 &  0 \\
                   0 &  0 &  1 &  0 & 1 &  1 &  0 \\
                   0 &  0 &  0 &  1 & 1 &  1 &  0 \\
                   0 & -1 & -1 & -1 & 1 &  1 & -2 \\
                   0 &  0 &  0 &  0 & 1 & -4 &  0 
                \end{array}\right)
\end{equation}

Here, the vector $v = (0,-1,-1,-1) = \frac{1}{2}(-1,-2,-2,-2) + \frac{1}{2}(1,0,0,0)$ arises through the blow-up of the unique singularity
in $\P_4^{(1,1,2,2,2)}$. Formula (\ref{hodge}) gives $h^{1,1}(M_1) = 2$ and furthermore from (\ref{vectorsM1}) we can deduce the following 
quantities
\begin{equation} \label{M1top}
  \begin{array}{ll}
  (1) & \mathcal{D}_1 = \Theta_1^2 (\Theta_1 - 2 \Theta_2) - 4 z_1 (4 \Theta_1 + 3) (4 \Theta_1 + 2) (4 \Theta_1 + 1) \\
   ~  & \mathcal{D}_2 = \Theta_2^2 - z_2 (2 \Theta_2 - \Theta_1 + 1) (2 \Theta_2 - \Theta_1) \\
   ~  & \Delta_{con} = -1 + 512 z_1 + 65536 z_1^2 (-1 + 4 z_2) \\
   ~  & \Delta_s = 1 - 4z_2 \\
  (2) & \kappa_{111} = 8, ~ \kappa_{112} = 4, ~ \kappa_{222} = \kappa_{221} = 0 \\
  (3) & \int_{M_1} c_2 J_1 = 56, ~ \int_{M_1} c_2 J_2 = 24 \\
  \end{array}
\end{equation}

\subsubsection{Solution at large radius}
\label{M1LargeRadius}

By the method of Frobenius we calculate the periods at large radius as solutions of the Picard-Fuchs system. This allows us to deduce the mirror map
from (\ref{defMirrorMap})
\begin{equation}
  \begin{array}{ccc}
    2 \pi i t_1(z_1, z_2) & = & \log(z_1) + 104 z_1 + 9780 z_1^2 -z_2 + 48 z_1 z_2 
                              - \frac{3}{2} z_2^2 + \cO(z^3), \\
    2 \pi i t_2(z_1, z_2) & = & \log(z_2) + 48 z_1 + 6408 z_1^2 + 2 z_2 - 96 z_1 z_2 + 3 z_2^2 + \cO(z^3).
  \end{array}
\end{equation}
These series can be inverted through introducing $q_i = e^{2 \pi i t_i}$ and one obtains
\begin{equation}
  \begin{array}{ccc}
    z_1(q_1,q_2) & = & q_1 -104 q_1^2 + 6444 q_1^3 + q_1 q_2 - 304 q_1^2 q_2 + \cO(q^4), \\
    z_2(q_1,q_2) & = & q_2 - 48 q_1 q_2 - 262 q_1^2 q_2 - 2 q_2^2 + 240 q_1 q_2^2 + 3 q_2^3 + \cO(q^4). 
  \end{array}
\end{equation}

The Yukawa-couplings can be deduced from the Picard-Fuchs system as well \cite{HSMS}. Using the classical intersection numbers for normalization
we obtain
\begin{eqnarray}
  C_{111} = \frac{-8}{z_1^3 \Delta_{con}}, & \quad &
  C_{112} = \frac{4 (256 z_1 - 1)}{z_1^2 z_2 \Delta_{con}}, \nonumber \\
  C_{122} = \frac{8 - 4096 z_1}{z_1 z_2 \Delta_s \Delta_{con}}, & \quad &
  C_{222} = \frac{-4 (1+4 z_2 - 256 z_1 (1+12 z_2))}{z_2^2 \Delta_s^2 \Delta_{con}}.
\end{eqnarray}

The Genus $0$ invariants can be expressed in terms of these through the expansion
\begin{equation} \label{gen0inf}
  K_{ijk} = \frac{1}{X_0^2} C_{ijk}(t_1, t_2) = \der_i \der_j \der_k F(t_1,t_2) 
          = \kappa_{ijk} + \sum_{d_1,d_2} \frac{n^0_{d_1, d_2} d_i d_j d_k}{1-\prod_{l=1}^2 q_l^{d_l}} \prod_{l=1}^2 q_l^{d_l}.
\end{equation}

In order to obtain the genus 1 free energy the holomorphic ambiguity has to be solved for. Using the ansatz (\ref{defF1b}) and as relevant boundary conditions
$\int c_2 J_i$ as well as the known genus one GV invariants we arrive at
\begin{equation}
  \cF^{(1)} = \log\left(\Delta_{con}^{-\frac{1}{12}} \Delta_s^{-\frac{5}{12}} \textrm{exp}\left[\frac{K}{2}(5-\frac{\chi}{12})\right] 
             \textrm{det}G^{-1}_{i\bar{j}} z_1^{-\frac{17}{6}} z_2^{-\frac{7}{8}}\right),
\end{equation}
and sending $\bar{t} \rightarrow \infty$ the following holomorphic limit
\begin{eqnarray}
  F^1(t_1,t_2) & = &-\frac{7}{3} \log(q_1) - \log(q_2) + \frac{160}{3} q_1 + \frac{2588}{3} q_1^2 + \frac{204928}{9} q_1^3 \nonumber \\
    ~          & ~ &+\frac{1}{3} q_2 + \frac{160}{3} q_1 q_2  + \frac{18056}{3} q_1^2 q_2 + \cO(q^4).
\end{eqnarray}

For higher genus calculations all propagators have to be obtained and $\cF^{(1)}$ has to be brought into the form (\ref{propF1}). Let us first concentrate
on the propagators, they are determined through the equations (\ref{propagators}) where the holomorphic ambiguities are solved for using symmetry properties.
For $S^{ij}$ independence on the index $k$ and symmetry in the indices $i$ and $j$ is enough to fix all $f^i_{lk}$:
\begin{eqnarray}
  f^1_{11} = -\frac{1}{z_1}, ~ f^1_{12} = - \frac{1}{4 z_2}, ~ f^1_{22} = 0, \nonumber \\
  f^2_{11} = 0, ~ f^2_{12} = \frac{1}{2z_1}, ~ f^2_{22} = - \frac{1}{z_2}. 
\end{eqnarray}
These choices lead to the following series expansions of the $S^{ij}$
\begin{eqnarray}
  S^{11} & = & - \frac{1}{16} z_1^2 + 16 z_1^3 + \frac{1}{4} z_1^2 z_2 - 152 z_1^3 z_2 + \cO(z^5), \nonumber \\
   S^{12} & = & \frac{1}{8} z_1 z_2 - 38 z_1^2 z_2 + 420 z_1^3 z_2 - \frac{1}{2} z_1 z_2^2 + 152 z_1^2 z_2^2 + \cO(z^5), \nonumber \\
  S^{22} & = & -\frac{1}{4} z_2^2 + 144 z_1^2 z_2^2 + z_2^3 + \cO(z^5). 
\end{eqnarray}
We find that the ambiguities appearing in the other propagators can be set to zero and obtain the following series expansions.
\begin{eqnarray}
  S^{1} & = & \frac{3}{2} z_1^2 - 105 z_1^3 - 12600 z_1^4 + 6 z_1^2 z_2 + 1548 z_1^3 z_2 + \cO(z^5), \nonumber  \\
  S^{2} & = & 3 z_1 z_2 + 282 z_1^2 z_2 + 33552 z_1^3 z_2 - 12 z_1 z_2^2 - 1128 z_1^2 z_2^2 + \cO(z^5), \nonumber \\
  S     & = & -18 z_1^2 - 2088 z_1^3 - 320328 z_1^4 - 144 z_1^2 z_2 - 60480 z_1^3 z_2 + \cO(z^5).
\end{eqnarray}
Next, we turn our attention to the truncation relations (\ref{trunc}). The ambiguities $h^{jk}_i$, $h^i_j$, $h_i$ and $h_{ij}$ are fixed through 
a series expansion of the holomorphic limit on both sides of (\ref{trunc})
\begin{eqnarray} \label{M1hamb}
  h^{11}_1 & = & - \frac{1}{32} z_1 (1-4 z_2 + 512 z_1 (11 z_2 -1)), ~ h^{12}_1 = \frac{1}{16} (704 z_1 - 1) z_2 (4 z_2-1), \nonumber \\
  h^{22}_1 & = & \frac{z_2^2 (4 z_2 - 1)}{8 z_1}, \nonumber \\
  h^{11}_2 & = & \frac{z_1^2 (1 + 4z_2-256 z_1 (1+15 z_2))}{64 z_2}, ~ h^{12}_2 = \frac{1}{32}z_1 (256 z_1-1) (1+4 z_2), \nonumber \\
  h^{22}_2 & = & \frac{1}{16}z_2 (1+4z_2 + 448 z_1 (4z_2-1)), \nonumber \\
  h^1_1   & = & 12 z_1 z_2 , ~ h_1^2 = 6 (1-4z_2)z_2, ~h_2^1 = 6 z_1^2, ~ h_2^2 = -12 z_1 z_2,
\end{eqnarray}
where all other ambiguities are either zero or follow by symmetry. Having obtained the truncation relations the process of direct integration
demands for the covariant derivative of $\cF^{(1)}$ in terms of the propagators, 
\begin{equation}
  D_i \cF^{(1)} = \frac{1}{2} C_{ijk} S^{jk} - \frac{1}{12}\Delta_{con}^{-1}\der_i \Delta_{con} - \frac{5}{12} \Delta_s^{-1} \der_i \Delta_s
                  + \der_i \log(z_1^{-\frac{31}{12}} z_2^{-\frac{7}{8}}).
\end{equation}

From here it is now straightforward to carry out the integration. First make an ansatz for $\cF^{(2)}$ as a polynomial of degree $3g-3$ in the generators
$\tilde{S}^{ij}$, $\tilde{S}^i$ and $\tilde S$ and evaluate the right hand side of the first equation in (\ref{Ftrunc}) by applying covariant derivatives
to $D_i \cF^{(1)}$ and using the truncation relations (\ref{trunc}). The right coefficients in $\cF^{(2)}$ follow then from comparison. In order to 
apply this procedure iteratively genus by genus the holomorphic ambiguity $f_g$ has to be fixed at each step. This is done as discussed in section (\ref{holamb})
by going to various boundary divisors in moduli space as will be described in the succeeding paragraphs.

\subsubsection{Solution at the Conifold locus}
\label{M1conifold}

We choose the monopole point of the Seiberg-Witten plane in order to carry out the conifold expansion. As was noted in section (\ref{SeibergWitten}) the 
correct coordinates at this point are 
\begin{equation}
  z_{c,1} = \frac{x^2 y}{(x-1)^2} - 1, ~ z_{c,2} = x - 1.
\end{equation}
Transforming the Picard-Fuchs system to these coordinates we find the following solutions (see also \cite{KlemmLuest})
\begin{eqnarray} \label{M1conperiods}
  \omega^c_0 & = & 1 + z_{c,2} - \frac{73}{64} z_{c,2}^2 - \frac{73}{192} z_{c,1} z_{c,2}^2 + \frac{7043}{4608} z_{c,2}^3 + \cO(z_c^4), \nonumber \\
  \omega^c_1 & = & z_{c,1} \sqrt{z_{c,2}} - \frac{15}{32} z_{c,1}^2 \sqrt{z_{c,2}} + \frac{315}{1024} z_{c,1}^3 \sqrt{z_{c,2}} 
                   + \frac{11}{64} z_{c,1}^2 z_{c,2}^{3/2} + \cO(z_c^5), \nonumber \\
  \omega^c_2 & = & z_{c,2} - \frac{35}{32} z_{c,2}^2 - \frac{35}{96} z_{c,1} z_{c,2}^2 + \frac{3325}{2304} z_{c,2}^3 + \cO(z_c^4),
\end{eqnarray}
where we have suppressed the dual logarithmic solutions. As mirror coordinates we take $t_{c,1} := \frac{\omega^c_1}{\omega^c_0}$ and
$t_{c,2} := \frac{\omega^c_2}{\omega^c_0}$. Relevant for the expansions around $z_{c,1}=z_{c,2}=0$ is the inverse mirror map given by
\begin{eqnarray}
  z_{c,1}(t_{c,1},t_{c,2}) & = & \frac{t_{c,1}}{\sqrt{t_{c,2}}} + \frac{15}{32} \frac{t_{c,1}^2}{t_{c,2}} 
                                 - \frac{3}{64} t_{c,1} \sqrt{t_{c,2}} + \frac{135}{1024} \frac{t_{c,1}^3}{t_{c,2}^{3/2}} 
                                 + \frac{1515}{65536} \frac{t_{c,1}^4}{t_{c,2}^2}
                                 - \frac{1223}{3072} t_{c,1}^2 + \cO(t_c^3), \nonumber \\
  z_{c,2}(t_{c,1},t_{c,2}) & = & t_{c,2} + \frac{67}{32} t_{c,2}^2 + \frac{35}{96} t_{c,1} t_{c,2}^{3/2} + \frac{175}{1024} t_{c,1}^2 t_{c,2}
                                 + \frac{18847}{4608} t_{c,2}^3 + \cO(t_c^4).
\end{eqnarray}
In order to perform the above inversion one has to define new variables $s_1 = \frac{t_{c,1}}{\sqrt{t_{c,2}}}$ and $s_2 = t_{c,2}$, calculate
$z_{c,i}(s_1, s_2)$ and then insert back $s_i(t_{c,i})$. The divisor $\{z_{c,2} = 0\}$ is normal to the conifold locus and $\{z_{c,1}=0\}$ 
is tangential. This means that $z_{c,1}$ parametrizes the normal direction to the conifold locus and $z_{c,2}$ the tangential one. Therefore
we expect $t_{c,1}$ to be appearing in inverse powers in the expansion of the free energies.

To obtain the free energies, all nonholomorphic generators appearing in the polynomial expansion of the $\cF^{(g)}$ have to be transformed into the 
$z_{c,i}$ coordinates. That is Yukawa couplings, the Christoffel symbols and the holomorphic ambiguities $f$ appearing in (\ref{propagators}) must
be transformed to the conifold coordinates. For the $f^i_{jk}$ this means
\begin{equation}
  f^i_{jk}(z_c) = \frac{\der z_{c,i}}{\der z_l} \left(\frac{\der^2 z_l}{\der z_{c,j} \der z_{c,k}}\right) 
                  + \frac{\der z_{ci}}{\der z_l} \frac{\der z_m}{\der z_{c,j}} \frac{\der z_n}{\der z_{c,k}} f^l_{mn}(z),
\end{equation} 
while the Yukawa-couplings have to be tensor transformed and the Christoffel symbols are obtained directly from the periods 
(\ref{M1conperiods}). 

We display our results for the $F^g_c$ up to genus 3:
\begin{eqnarray}
  F_c^1 & = & -\frac{1}{12} \log\left(\frac{t_{c,1}}{\sqrt{t_{c,2}}}\right) - \frac{29}{12} \log(t_{c,2}) 
              - \frac{137}{128}\frac{t_{c,1}}{\sqrt{t_{c,2}}} - \frac{9827}{768} t_{c,2} 
              + \frac{189}{8192} \frac{t_{c,1}^2}{t_{c,2}} + \cO(t_c^2), \nonumber \\
  F_c^2 & = & -\frac{1}{240 t_{c,1}^2} + \frac{155359}{589824} - \frac{550551 t_{c,1}^2}{33554432 t_{c,2}^2} - \frac{18321 t_{c,1}}{524288 t_{c,2}^{3/2}}
              + \frac{1067}{6144 t_{c,2}} + \cO(t_c^1), \nonumber \\ 
  {F_c}^3 & = & \frac{1}{1008 t_{c,1}^4} + \frac{788437361}{21743271936} + \cO(t_c^1) 
\end{eqnarray}

\subsubsection{Solution at the strong coupling locus}
\label{M1StrongCoupling}

We expand around the point of intersection of the divisors $C_1$ and $D_{0,-1}$. The right coordinates are
\begin{equation}
  z_{s,1} = \Delta_s^{\frac{1}{2}}, ~ z_{s,2} = x_1.
\end{equation}

The Picard-Fuchs system at this point contains among its solutions a one logarithmic in $x_1$ which is a 
continuation of the logarithmic solution already present at large radius.
\begin{eqnarray}
  \omega^s_0 & = & 1 + \frac{3}{32} z_{s,2} + \frac{945}{16384} z_2^2 + \frac{28875}{524288} z_2^3 + \cO(z_s^4), \nonumber \\
  \omega^s_1 & = & z_{s,1} + \frac{z_{s,1}^3}{3} + \frac{1}{32} z_{s,1}^3 z_{s,2} + \cO(z_s^5), \nonumber \\
  \omega^s_2 & = & \omega^s_0 \log(z_{s,2}) - \frac{1}{2} z_{s,1}^2 + \frac{z_{s,2}}{2} + \frac{2853}{8192} z_{s,2}^2 - \frac{3}{64} z_{s,1}^2 z_{s,2} \nonumber \\
     ~       & ~ & + \frac{273425}{786432} z_{s,2}^3 + \cO(z_s^4).
\end{eqnarray}
The mirror map is deduced from the quotients by $\omega^s_0$: $t_{s,1} := \frac{\omega^s_1}{\omega^s_0}$ , $t_{s,2} := \frac{\omega^s_2}{\omega^s_0}$.
Building the inverse we arrive at 
\begin{eqnarray}
  z_{s,1}(t_{s,1},q_{s,2}) & = & t_{s,1} + \frac{3}{32} q_{s,2} t_{s,1} + \frac{177}{16384} q_{s,2}^2 t_{s,1} - \frac{t_{s,1}^3}{3} + \cO(t_c^4), \nonumber \\
  z_{s,2}(t_{s,1},q_{s,2}) & = & q_{s,2} - \frac{q_{s,2}^2}{2} + \frac{603}{8192} q_{s,2}^3 + \frac{1}{2} q_{s,2} t_{s,1}^2 + \cO(t_c^4).
\end{eqnarray}

Transforming the Yukawa couplings, the Christoffel symbols and the holomorphic ambiguities $f$ to the strong coupling coordinates we obtain the 
propagators at this point. Tensor transforming the propagators to $z_{i}$ coordinates and substituting in all holomorphic quantities in the $\cF^{(g)}$
$z_i \rightarrow z_i(z_{s,1},z_{s,2})$ we obtain the $\cF^{(g)}_s$. In the holomorphic limit this gives the following genus 1,2 and 3 expansions.
\begin{eqnarray}
  F_s^1 & = & -\frac{1}{3} \log(t_{s,1}) - \frac{7}{3} t_{s,2} + \frac{5}{24} q_{s,2}
              + \frac{121}{4096} q_{s,2}^2 - \frac{t_{s,1}}{18} + \frac{5}{48} q_{s,2} t_{s,1} \nonumber \\
  ~     & ~ & + \frac{t_{s,2}^2}{540} + \cO(t_s^3), \nonumber \\
  F_s^2 & = & \frac{1}{240 t_{s,1}^2} - \frac{11}{360} + \frac{q_{s,2}}{1280} - \frac{1057}{2097152} \frac{q_{s,2}^3}{t_{s,1}^2} + \cO(t_s^2) , \nonumber \\
  F_s^3 & = & \frac{1}{4032 t_{s,1}^4} + \frac{11}{90720} - \frac{q_{s,2}}{16128} + \frac{18805}{132120576} \frac{q_{s,1}^3}{t_{s,1}^2} \nonumber \\
    ~   & ~ &  - \frac{62210349}{549755813888} \frac{q_{s,2}^5}{t_{s,1}^4} + \cO(t_s^2).
\end{eqnarray}
It is important to note here that the variable $t_{s,1}$ is not the true variable characterizing the size of the shrinking cycle. The true
size is given by the rescaling $t_{s,1} \rightarrow 2 t_{s,1}$. The factor of $4$ which then arises in the expansions in comparison to the conifold case
is exactly the difference between hyper- and vectormultiplets calculated in section (\ref{StrongCoupling}). However, note that we have only displayed results up to genus $3$. For genus $4$ we find that there is no simple gap structure and therefore the boundary conditions at this point remain unclear. Also, we find that for genus $4$ the parametrization of the ambiguity changes slightly as the numerator of the strong coupling discriminant has to be parametrized to a higher degree
\begin{equation} \label{sf4}
  f_g = \ldots + \frac{\sum_{|I| \leq 2g-2} c^s_I z^I}{\Delta_s^{g-1}} + \ldots .
\end{equation}

As in this modified ansatz some of the parameters lead to contributions of nonregular terms at the Gepner point they will be solved for by the regularity condition at that point.

The strong coupling divisor provides us with a further boundary condition to check the consistency of our approach. At this locus there is an extremal transition to the 1-parameter complete intersection $\P^5[4,2]$. The free energy expansions for $\P^5[4,2]$ should come out naturally from the expansion at large radius of our 2-parameter
model by setting $q_2 = 1$ which corresponds to shrinking the size of the $\P^1$ base to zero
\footnote{In the general case the transition is obtained by replacing $(2g-2)\left(\begin{array}{c}N\\2\end{array}\right)$ two spheres by
$(2g-2)\left(\begin{array}{c}N\\2\end{array}\right)$ three spheres, where $g$ is the genus of the singular curve and $N-1$ is the rank of the gauge group. 
This results in a change of Hodge numbers given by $h^{1,1} \mapsto h^{1,1} - (N-1),~~ 
h^{2,1} \mapsto h^{2,1} + (2g-2) \left(\begin{array}{c}N\\2\end{array}\right) - (N-1)$.}. 
Therefore the Gopakumar-Vafa invariants of $\P^5[4,2]$
should be equal to the sum over the second degree of the invariants of $M_1$, i.e. $n_k^{(g)} = \sum_{d_2} n^{(g)}_{d_1 d_2}$. This is indeed true
as can be checked from the tables in appendix \ref{GVInvariants}.

\subsubsection{Solution at the Gepner point}
\label{M1Gepner}

Here we expand around the point of intersection of the divisors $C_0$ and $D_{-1,0}$ which corresponds to $\psi = \phi = 0$. This gives the 
variables
\begin{equation} \label{ovars}
  z_{o,1} = \psi = \frac{1}{(z_2 z_1^2)^{\frac{1}{8}}}, ~ z_{o,2} = \phi = \frac{1}{z_2^{\frac{1}{2}}},
\end{equation}
where we have chosen the subscript $o$ as the Gepner point is a Landau-Ginzburg orbifold. The transformed Picard-Fuchs system admits the solutions
\begin{eqnarray}
  \omega^o_0 & = & z_{o,1} + \frac{1}{32} z_{o,1} z_{o,2}^2 + \frac{27}{2048} z_{o,1} z_{o,2}^4 + \cO(z_o^6), \nonumber \\
  \omega^o_1 & = & z_{o,1} z_{o,2} + \frac{25}{96} z_{o,1} z_{o,2}^3 + \frac{z_{o,1}^5}{6} + \cO(z_o^6), \nonumber \\
  \omega^o_2 & = & z_{o,1}^2 + \frac{1}{8} z_{o,1}^2 z_{o,2}^2 + \cO(z_o^6), \nonumber \\
  \omega^o_3 & = & z_{o,1}^2 z_{o,2} + \frac{3}{8} z_{o,1}^2 z_{o,2}^3 + \cO(z_o^6), \nonumber \\
  \omega^o_4 & = & z_{o,1}^3 + \frac{9}{32} z_{o,1}^3 z_{o,2}^2 + \cO(z_o^6), 
\end{eqnarray}
where we have omitted the solution corresponding to the 6th period. From the above we extract the mirror map $t_{1,o} := \frac{\omega^o_1}{\omega^o_0}$,
$t_{2,o} := \frac{\omega^o_2}{\omega^o_0}$ and its inverse
\begin{eqnarray} \label{omirror}
  z_{o,1}(t_{o,1},t_{o,2}) & = & t_{o,1} - \frac{3}{32} t_{o,1} t_{o,2}^2 + \frac{17}{6144} t_{o,1} t_{o,2}^4 + \cO(t_o^6), \nonumber \\
  z_{o,2}(t_{o,1},t_{o,2}) & = & t_{o,2} - \frac{11}{48} t_{o,2}^3 - \frac{1}{6} t_{o,1}^4 + \frac{31}{768} t_{o,2}^5 + \cO(t_o^6).
\end{eqnarray}
The genus 0 Prepotential can be extracted from the periods $\omega^o_3$ and $\omega^o_4$ through the special geometry relation
\begin{equation}
  \omega^o_3 = \omega^o_0 \left(\frac{\der}{\der t_{o,1}} F_o(t_{o,1},t_{o,2})\right), ~ 
  \omega^o_4 = \omega^o_0 \left(\frac{\der}{\der t_{o,2}} F_o(t_{o,1},t_{o,2})\right),
\end{equation}
yielding
\begin{equation}
  F_o(t_{o,1},t_{o,2}) = t_{o,1}^2 t_{o,2} + \frac{1}{48} t_{o,1}^2 t_{o,2}^3 + \frac{t_{o,1}^6}{30} + \cO(t_o^7).
\end{equation}

The $F_o^g(t_{o,1},t_{o,2})$ can be calculated in the same way as in the case of the other boundary divisors and we find that
as expected the correct GV-invariants are produced once we require the free energies to be regular at the orbifold point.
This way also all polynomial ambiguities $a_I$ are fixed uniquely. Let us delve a bit more into the details at this point
as the question of fixing the $a_I$ is ultimately related to the question of integrability. We observe that the anomaly part of the $F^g_o$ always comes
with a pole $\frac{1}{z_{o,1}^{2g-2}} + \ldots$ and that no poles of type $\frac{1}{z_{o,2}}$ appear. This can be traced
back to the expansion of propagators and Yukawa couplings around this point. Therefore, we are able to constrain the parametrization
of the ambiguity in the following way: The coefficients $a_{i_1,i_2}$ will be nonvanishing only for indices $i_2 \leq i_1/2$. The reason
can be found in the relations inverse to (\ref{ovars})
\begin{equation}
  z_1 = - \frac{z_{o,2}}{256 z_{o,4}^4}, \quad z_2 = \frac{1}{4 z_{o,2}^2}.
\end{equation}

One can see that in order to avoid poles in the second variable $z_{o,2}$ one has to multiply each power of $z_2$ with a power of $z_1$ 
which is at least two times larger. We find that at genus $2$ the regularity condition is enough to fix all $a_I$'s without exception. 
On the other hand the calculation at genus $3$ yields that regularity is only strong enough to fix all but one of the $a_I$, namely 
$a_{1,0}$. However, this single parameter is already solved for by the knowledge of the fiber invariants extracted from the weak coupling divisor. 
At genus $4$ we find that two parameters remain unfixed after having imposed regularity, namely $a_{1,0}$ and $a_{2,0}$. Again these two 
will be ultimately solved for by the knowledge of the fibre invariants once all other parameters are fixed. This procedure will carry on up to
genus infinity once the parameters related to the strong coupling divisor and the conifold divisor can be fixed at each genus by appropriate 
boundary conditions.

We display our results for the higher $F_o^g(t_{o,1},t_{o,2})$ ($g \geq 1$) :
\begin{eqnarray}
  F_o^1(t_{o,1},t_{o,2}) & = & -\frac{3}{16} t_{o,2}^2 + \frac{7}{384} t_{o,2}^4 - \frac{1}{16} t_{o,1}^4 t_{o,2} + \frac{43}{184320} t_{o,2}^6 - \frac{5}{256} t_{o,1}^4 t_{o,2}^3\nonumber \\
        ~               & ~ & -\frac{5}{252} t_{o,1}^8 + \frac{2237}{20643840} t_{o,2}^8 - \frac{35}{6144} t_{o,1}^4 t_{o,2}^5 - \frac{7}{1280} t_{o,1}^8 t_{o,2}^2 \nonumber \\
        ~               & ~ & + \frac{40603}{2972712960} t_{o,2}^{10} + \cO(t_o^{11}), \nonumber \\
  F_o^2(t_{o,1},t_{o,2}) & = & \frac{113}{7680} t_{o,1}^2 - \frac{377}{122880} t_{o,1}^2 t_{o,2}^2 + \frac{363}{655360} t_{o,1}^2 t_{o,2}^4 \nonumber \\
        ~               & ~ & + \frac{59}{307200} t_{o,1}^6 t_{o,2} - \frac{153361}{707788800} t_{o,1}^2 t_{o,2}^6 + \frac{39041}{44236800} t_{o,1}^6 t_{o,2}^3 \nonumber \\
        ~               & ~ & - \frac{143}{3440640} t_{o,1}^{10} - \frac{10379101}{158544691200} t_{o,1}^2 t_{o,2}^8 + \cO(t_o^{11}), \nonumber 
\end{eqnarray}

\begin{eqnarray}
  F_o^3(t_{o,1},t_{o,2}) & = & - \frac{61}{3932160} t_{o,2} + \frac{29}{9437184} t_{o,2}^3 + \frac{875}{4718592} t_{o,1}^4 - \frac{581}{603979776} t_{o,2}^5 \nonumber \\
        ~               & ~ & + \frac{1585}{44040192} t_{o,1}^4 t_{o,2}^2 - \frac{236533}{1014686023680} t_{o,2}^7 + \frac{1221673}{12683575296} t_{o,1}^4 t_{o,2}^4 \nonumber \\
        ~               & ~ & - \frac{43439}{1056964608} t_{o,1}^8 t_{o,2} - \frac{6903751}{73057393704960} t_{o,2}^9 + \frac{20689415}{304405807104} t_{o,1}^4 t_{o,2}^6 \nonumber \\
        ~               & ~ & + \cO(t_o^{11}).
\end{eqnarray}

\subsection{$M_2 = \P_4^{(1,1,2,2,6)}[12]$}
\label{M2}

The toric data of this Calabi-Yau is summarized in the following matrix.

\begin{equation}  \label{vectorsM2}
  (V|L) = \left(\begin{array}{ccccc|cc}
                  -1 & -2 & -2 & -6 & 1 &  0 &  1 \\
                   1 &  0 &  0 &  0 & 1 &  0 &  1 \\
                   0 &  1 &  0 &  0 & 1 &  1 &  0 \\
                   0 &  0 &  1 &  0 & 1 &  1 &  0 \\
                   0 &  0 &  0 &  1 & 1 &  3 &  0 \\
                   0 & -1 & -1 & -3 & 1 &  1 & -2\\
                   0 &  0 &  0 &  0 & 1 & -6 & 0 
                \end{array}\right)
\end{equation}

The vector introduced through the blow-up of the unique singularity in $\P_4^{(1,1,2,2,2)}$ is 
$v = (0,-1,-1,-3) = \frac{1}{2}(-1,-2,-2,-6) + \frac{1}{2}(1,0,0,0)$. Again we obtain from formula (\ref{hodge}) $h^{1,1}(M_1) = 2$ and from (\ref{vectorsM2})
\begin{equation} \label{M2top}
  \begin{array}{ll}
  (1) & \mathcal{D}_1 = \Theta_1^2 (\Theta_1 - 2 \Theta_2) - 8 z_1 (6 \Theta_1 + 5) (6 \Theta_1 + 3) (6 \Theta_1 + 1) \\
   ~  & \mathcal{D}_2 = \Theta_2^2 - z_2 (2 \Theta_2 - \Theta_1 + 1) (2 \Theta_2 - \Theta_1) \\
   ~  & \Delta_{con} = 1 - 3456 z_1 -2985984 z_1^2 (-1 + 4 z_2) \\
   ~  & \Delta_s = 1 - 4z_2 \\
  (2) & \kappa_{111} = 4, ~ \kappa_{112} = 2, ~ \kappa_{222} = \kappa_{221} = 0 \\
  (3) & \int_{M_1} c_2 J_1 = 52, ~ \int_{M_1} c_2 J_2 = 24 \\
  \end{array}
\end{equation}

\subsubsection{Solution at large radius}
\label{M2LargeRadius}

As a first step we calculate the periods at large radius as solutions of the Picard-Fuchs system. The corresponding mirror map 
from (\ref{defMirrorMap}) is
\begin{equation}
  \begin{array}{ccc}
    2 \pi i t_1(z_1, z_2) & = & \log(z_1) + 744 z_1 + 473652 z_1^2 - z_2 + 240 z_1 z_2 - \frac{3}{2} z_2^2 + \cO(z^3), \\
    2 \pi i t_2(z_1, z_2) & = & \log(z_2) + 240 z_1 + 220680 z_1^2 + 2 z_2 - 480 z_1 z_2 + 3 z_2^2 + \cO(z^3).
  \end{array}
\end{equation}
Inverting these series we obtain ($q_i = e^{2 \pi i t_i}$) 
\begin{equation}
  \begin{array}{ccc}
    z_1(q_1,q_2) & = & q_1 -744 q_1^2 + 356652 q_1^3 + q_1 q_2 - 1968 q_1^2 q_2 + \cO(q^4), \\
    z_2(q_1,q_2) & = & q_2 - 240 q_1 q_2 - 13320 q_1^2 q_2 - 2 q_2^2 + 1200 q_1 q_2^2 + 3 q_2^3 + \cO(q^4). 
  \end{array}
\end{equation}

The Yukawa-couplings are extracted from the Picard-Fuchs system and for normalization we use the classical intersection numbers giving
\begin{eqnarray}
  C_{111} = \frac{4}{z_1^3 \Delta_{con}}, & \quad &
  C_{112} = \frac{2-3456 z_1}{z_1^2 z_2 \Delta_{con}}, \nonumber \\
  C_{122} = \frac{13824 z_1 - 4}{z_1 z_2 \Delta_s \Delta_{con}}, & \quad &
  C_{222} = \frac{2-8 z_2 - 3456 z_1 (1 + 12 z_2)}{z_2^2 \Delta_s^2 \Delta_{con}}.
\end{eqnarray}

The Genus $0$ invariants can be obtained in terms of these through the expansion (\ref{gen0inf}).

The genus 1 free energy is fixed by the known GV-invariants and boundary conditions 
\begin{equation}
  \cF^{(1)} = \log\left(\Delta_{con}^{-\frac{1}{12}} \Delta_s^{-\frac{1}{3}} \textrm{exp}\left[\frac{K}{2}(5-\frac{\chi}{12})\right] 
             \textrm{det}G^{-1}_{i\bar{j}} z_1^{-\frac{8}{3}} z_2^{-\frac{3}{2}}\right),
\end{equation}
giving for $\bar{t} \rightarrow \infty$ the series
\begin{eqnarray}\label{M2F1}
  F^1(t_1,t_2) & = & \frac{13}{6} \log(q_1) + \log(q_2) - 208 q_1 - 18258 q_1^2 - \frac{5261632}{3}q_1^3 - \frac{q_2}{6} \nonumber \\
    ~          & ~ &  - 208 q_1 q_2 - 162252 q_1^2 q_2 - \frac{q_2^2}{12} - \frac{q_2^3}{18}+ \cO(q^4).
\end{eqnarray}

Higher genus calculations require the expansion of propagators and the polynomial form for $\cD_i \cF^{(1)}$ (\ref{propF1}). The ambiguities in the
propagators are fixed through symmetry considerations as explained in (\ref{M1LargeRadius}). We obtain
\begin{eqnarray}
  f^1_{11} = -\frac{1}{z_1}, ~ f^1_{12} = - \frac{1}{4 z_2}, ~ f^1_{22} = 0, \nonumber \\
  f^2_{11} = 0, ~ f^2_{12} = \frac{1}{2z_1}, ~ f^2_{22} = - \frac{1}{z_2},
\end{eqnarray}
while the ambiguities appearing in the propagators $S^i$ and $S$ can all be set to zero.
These choices lead to the following series expansions
\begin{eqnarray}
  S^{11} & = & - \frac{1}{8} z_1^2 + 216 z_1^3 + \frac{1}{2} z_1^2 z_2 - 1968 z_1^3 z_2 + \cO(z^5), \nonumber \\
   S^{12} & = & \frac{1}{4} z_1 z_2 - 492 z_1^2 z_2 + 27720 z_1^3 z_2 - z_1 z_2^2 + 1968 z_1^2 z_2^2 + \cO(z^5), \nonumber \\
  S^{22} & = & -\frac{1}{2} z_2^2 + 7200 z_1^2 z_2^2 + 2 z_2^3 + \cO(z^5), \nonumber \\
  S^{1} & = & 15 z_1^2 - 6930 z_1^3 - 5710320 z_1^4 + 60 z_1^2 z_2 + 97560 z_1^3 Z+ \cO(z^5), \nonumber  \\
  S^{2} & = & 30 z_1 z_2 + 17460 z_1^2 z_2 + 14315040 z_1^3 z_2 - 120 z_1 z_2^2 - 69840 z_1^2 z_2^2 + \cO(z^5), \nonumber \\
  S     & = & -900 z_1^2 - 723600 z_1^3 - 757242000 z_1^4 - 7200 z_1^2 z_2 \nonumber \\
  ~     & ~ & - 19958400 z_1^3 z_2 + \cO(z^5). 
\end{eqnarray}

In order to obtain the truncation relations (\ref{trunc}) the ambiguities $h^{jk}_i$, $h^i_j$, $h_i$ and $h_{ij}$ are fixed by taking 
the holomorphic limit of (\ref{trunc}).
\begin{eqnarray} \label{M2hamb}
  h^{11}_1 & = & - \frac{1}{16} z_1 (1-4 z_2 + 384 z_1 (92 z_2 - 9)), ~  h^{21}_1 =  \frac{1}{8}(4416 z_1 - 1)z_2 (4 z_2 -1 ), \nonumber \\
  h^{22}_1 & = & \frac{z_2^2 (4 z_2 -1)}{4 z_1}, \nonumber \\
  h^{11}_2 & = & \frac{z_1^2 (1 + 4 z_2 - 192 z_1 (9 + 128 z_2))}{32 z_2}, ~ h^{12}_2 = \frac{1}{16} z_1 (1728 z_1 - 1)(1 + 4 z_2), \nonumber \\
  h^{22}_2 & = & \frac{1}{8} z_2 (1 + 4 z_2 + 2688 z_1 (4 z_2 -1)), \nonumber \\
  h^1_1   & = & 120 z_1 z_2, ~h_2^1 = 60 z_1^2, h_1^2 = 60 (1-4z_2) z_2, ~ h_2^2 = -120 z_1 z_2,
\end{eqnarray}
where all other ambiguities are either zero or follow by symmetry. The dependence of the covariant derivative of $\cF^{(1)}$ on holomorphic 
and non-holomorphic generators is obtained from comparison with the expansion (\ref{M2F1}). 
\begin{equation}
  D_i \cF^{(1)} = \frac{1}{2} C_{ijk} S^{jk} - \frac{1}{12}\Delta_{con}^{-1}\der_i \Delta_{con} - \frac{1}{3} \Delta_s^{-1} \der_i \Delta_s
                  + \der_i \log(z_1^{-\frac{29}{12}} z_2^{-\frac{7}{8}}).
\end{equation}

The direct integration procedure is now straight forward as discussed at the end of section (\ref{M1LargeRadius}). In the following sections
we describe the solutions of the Picard-Fuchs system at the various boundary divisors.

\subsubsection{Solution at the Conifold locus}
\label{M2conifold}

We choose the coordinates introduced in (\ref{SeibergWitten})
\begin{equation}
  z_{c,1} = \frac{x^2 y}{(x-1)^2} - 1, ~ z_{c,2} = x - 1.
\end{equation}
Then the transformed Picard-Fuchs system admits the following set of solutions (see also \cite{KlemmLuest})
\begin{eqnarray} \label{M2conperiods}
  \omega^c_0 & = & 1 + z_{c,2} - \frac{53}{48} z_{c,2}^2 - \frac{53}{144} z_{c,1} z_{c,2}^2 + \frac{11371}{7776} z_{c,2}^3 + \cO(z_c^4), \nonumber \\
  \omega^c_1 & = & z_{c,1}\sqrt{z_{c,2}} - \frac{15}{32}z_{c,1}^2 \sqrt{z_{c,2}} + \cO(z_c^4), \nonumber \\
  \omega^c_2 & = & z_{c,2} - \frac{77}{72} z_{c,2}^2 - \frac{77}{216} z_{c,1} z_{c,2}^2 + \frac{4081}{2916} z_{c,2}^3 + \cO(z_c^4),
\end{eqnarray}
where we have suppressed the dual logarithmic solutions. As mirror coordinates we take $t_{c,1} := \frac{\omega^c_1}{\omega^c_0}$ and
$t_{c,2} := \frac{\omega^c_2}{\omega^c_0}$. Inverting the mirror map gives
\begin{eqnarray}
  z_{c,1}(t_{c,1},t_{c,2}) & = & \frac{t_{c,1}}{\sqrt{t_{c,2}}} + \frac{15}{32} \frac{t_{c,1}^2}{t_{c,2}} 
                                + \frac{1515}{65536}\frac{t_{c,1}^4}{t_{c,2}^2} - \frac{2561}{6912} t_{c,1}^2
                                + \frac{5 (243 t_{c,1}^3 - 64 t_{c,1} t_{c,2}^2)}{9216 t_{c,2}^{\frac{3}{2}}} + \cO(t_c^3), \nonumber \\
  z_{c,2}(t_{c,1},t_{c,2}) & = & t_{c,2} + \frac{149}{72} t_{c,2}^2 + \frac{77}{216} t_{c,1} t_{c,2}^{\frac{3}{2}}
                               + \frac{385}{2304} t_{c,1}^2 t_{c,2} + \frac{93127}{23328} t_{c,2}^3 + \cO(t_c^4).
\end{eqnarray}

After transforming all nonholomorphic generators appearing in the polynomial expansion of the $\cF^{(g)}$ into the 
$z_{c,i}$ coordinates, the $\cF^{(g)}_c$ are obtained by sending 
$z_i \rightarrow z_i(z_{c,1},z_{c,2})$.  We display our results up to genus 3:
\begin{eqnarray}
  F_c^1 & = & -\frac{1}{12}\log\left(\frac{t_{c,1}}{\sqrt{t_{c,2}}}\right) - \frac{29}{12}\log\left(t_{c,2}\right)
              - \frac{137}{128}\frac{t_{c,1}}{\sqrt{t_{c,2}}}-\frac{7255}{576} t_{c,2}
              + \frac{189}{8192}\frac{t_{c,1}^2}{t_{c,2}} + \cO(t_c^2), \nonumber \\
  F_c^2 & = & -\frac{1}{120 t_{c,1}^2} + \frac{1468157}{3317760} + \frac{1067}{3072} \frac{1}{t_{c,2}} - \frac{18321}{262144}\frac{1}{t_{c,2}^{3/2}}
              - \frac{550551}{16777216}\frac{t_{c,1}^2}{t_{c,2}^2}  + \cO(t_c^1), \nonumber \\ 
  {F_c}^3 & = & \frac{1}{252 t_{c,1}^4} + \frac{17978057749}{137594142720} + \cO(t_c^1) 
\end{eqnarray}
Here, $t_{c,1}$ is not the true size of the shrinking cycle, it has to be rescaled $t_{c,1} \rightarrow \frac{t_{c,1}}{\sqrt{2}}$.

\subsubsection{Solution at the strong coupling locus}
\label{M2StrongCoupling}

We expand around the point of intersection of the divisors $C_1$ and $D_{0,-1}$ with the following coordinate choice
\begin{equation}
  z_{s,1} = \Delta_s^{\frac{1}{2}}, ~ z_{s,2} = x_1.
\end{equation}

We present the three solutions of the Picard-Fuchs system which are relevant for the mirror map. As in the case of the
previous model there is a logarithmic solution coming from the logarithmic solution at large radius.
\begin{eqnarray}
  \omega^s_0 & = & 1 + \frac{5}{72} z_{s,2} + \frac{385}{9216} z_{s,2}^2 + \frac{2127125}{53747712} z_{s,s}^3 + \cO(z_s^4), \nonumber \\
  \omega^s_1 & = & z_{s,1} + \frac{z_{s,1}^3}{3} + \frac{5}{216} z_{s,1}^3 z_{s,2} + \cO(z_s^4), \nonumber \\
  \omega^s_2 & = & \omega^s_0 \log(z_{s,2}) + \frac{z_{s,2}}{2} + \frac{9166 z_{s,2}^2 - 13824 z_{s,1}^2}{27648}  \nonumber \\
     ~       & ~ & - \frac{5(373248 z_{s,1}^2 z_{s,2} - 3495750 z_{s,2}^3)}{53747712}.
\end{eqnarray}
The mirror map is deduced from the quotients by $\omega^s_0$: $t_{s,1} := \frac{\omega^s_1}{\omega^s_0}$ , $t_{s,2} := \frac{\omega^s_2}{\omega^s_0}$.
Building the inverse we arrive at
\begin{eqnarray}
  z_{s,1}(t_{s,1},q_{s,2}) & = & t_{s,1} + \frac{5}{72} q_{s,2} t_{s,1} + \frac{65}{9216} q_{s,2}^2 t_{s,1} - \frac{t_{s,1}^3}{3} + \cO(t_c^4), \nonumber \\
  z_{s,2}(t_{s,1},q_{s,2}) & = & q_{s,2} - \frac{q_{s,2}^2}{2} + \frac{1081}{13824} q_{s,2}^3 + \frac{1}{2} q_{s,2} t_{s,1}^2 + \cO(t_c^4).
\end{eqnarray}
The period describing the vanishing $\P^1$ is $\omega^s_1$. Therefore the right coordinate appearing in inverse powers in the $F^{(g)}_s$ should be
$t_{s,1}$. 

After having transformed all holomorphic and non-holomorphic quantities to the strong coupling point we list the genus 1,2 and 3 expansions of the
free energy.
\begin{eqnarray} \label{M2sExpansion}
  F_s^1 & = & -\frac{1}{6}\log t_{s,1} - \frac{13}{6} \log(q_{s,2}) + \frac{157}{432}q_{s,2} 
              + \frac{16367}{497664} q_{s,2}^2 - \frac{t_{s,1}^2}{36} + \cO(t_s^4), \nonumber \\ 
  F_s^2 & = & \frac{1}{480 t_{s,1}^2} - \frac{2}{45} + \cO(q_s^1) , \nonumber \\
  F_s^3 & = & \frac{1}{8064 t_{s,1}^4} + \frac{1}{5670} - \frac{5}{217728} q_{s,2} - \frac{64100947405}{1711891286065152} \frac{q_{s,2}^5}{t_{s,1}^4}\nonumber \\
    ~   & ~ &  + \frac{3869975}{81266540544} \frac{q_{s,2}^3}{t_{s,1}^2}+ \cO(t_s^1).
\end{eqnarray}
Again we stress that the variable $t_{s,1}$ is not the true variable characterizing the size of the shrinking cycle. The right scaling is given by
$t_{s,1} \rightarrow 2 t_{s,1}$ and the factor of $2$ which then arises in the numerator of the leading singularity 
is exactly the difference between hyper- and vectormultiplets calculated in section (\ref{StrongCoupling}). At this point we have to stress that there no simple gap as apparent in the equations (\ref{M2sExpansion}) and the correct boundary conditions remain unclear.

\subsubsection{Solution at the Gepner point}
\label{M2Gepner}

Expanding around the point of intersection of the divisors $C_0$ and $D_{-1,0}$ corresponds to to the choice
\begin{equation}
  z_{o,1} = \psi = \frac{1}{(z_2 z_1^2)^{\frac{1}{12}}}, ~ z_{o,2} = \phi = \frac{1}{z_2^{\frac{1}{2}}},
\end{equation}
where we have chosen the subscript $o$ as the Gepner point is a Landau-Ginzburg orbifold. The transformed Picard-Fuchs system admits the solutions
\begin{eqnarray}
  \omega^o_0 & = & z_{o,1} + \frac{1}{72} z_{o,1} z_{o,2}^2 + \frac{169}{31104} z_{o,1} z_{o,2}^4 + \cO(z_o^6), \nonumber \\
  \omega^o_1 & = & z_{o,1} z_{o,2} + \frac{49}{216} z_{o,1} z_{o,2}^3 + \cO(z_o^6), \nonumber \\
  \omega^o_2 & = & z_{o,1}^3 + \frac{1}{8} z_{o,1}^3 z_{o,2}^2  + \cO(z_o^6), \nonumber \\
  \omega^o_3 & = & z_{o,1}^3 z_{o,2} + \frac{3}{8} z_{o,1}^3 z_{o,2}^3 + \frac{147}{640} z_{o,1}^3 z_{o,2}^5 + \cO(z_o^9), \nonumber \\
  \omega^o_4 & = & \frac{z_{o,1}^5}{2} + \frac{25}{144} z_{o,1}^5 z_{o,2}^2 + \cO(z_o^8), 
\end{eqnarray}
where we have omitted the solution corresponding to the 6th period. From the above we extract the mirror map $t_{1,o} := \frac{\omega^o_1}{\omega^o_0}$,
$t_{2,o} := \frac{\omega^o_2}{\omega^o_0}$ and its inverse
\begin{eqnarray}
  z_{o,1}(t_{o,1},t_{o,2}) & = & \sqrt{t_{o,1}} - \frac{5}{256} t_{o,1}^{7/2} t_{o,2} - \frac{1}{18} \sqrt{t_{o,1}} t_{o,2}^2  
                               - \frac{1}{1296} \sqrt{t_{o,1}} t_{o,2}^4 + \cO(t_o^6), \nonumber \\
  z_{o,2}(t_{o,1},t_{o,2}) & = & t_{o,2} - \frac{1}{8} t_{o,1}^3 t_{o,2} - \frac{23}{108} t_{o,2}^3 - \frac{1}{8} t_{o,1}^3 + \frac{199}{6480} t_{o,2}^5 + \cO(t_o^6).
\end{eqnarray}
The genus 0 Prepotential can be extracted from the periods $\omega^o_3$ and $\omega^o_4$ through the special geometry relation
\begin{equation}
  \omega^o_3 = \omega^o_0 \left(\frac{\der}{\der t_{o,1}} F_o(t_{o,1},t_{o,2})\right), ~ 
  \omega^o_4 = \omega^o_0 \left(\frac{\der}{\der t_{o,2}} F_o(t_{o,1},t_{o,2})\right),
\end{equation}
yielding
\begin{equation}
  F_o(t_{o,1},t_{o,2}) = \frac{1}{2} t_{o,1}^2 t_{o,2} + \frac{1}{32} t_{o,1}^5 + \frac{1}{54} t_{o,1}^2 t_{o,2}^3 + \frac{5}{384} t_{o,1}^5 t_{o,2}^2
                        + \frac{7}{3240} t_{o,1}^2 t_{o,2}^5 + \cO(t_o^9).
\end{equation}

The $F_o^g(t_{o,1},t_{o,2})$ can be calculated in the same way as in the case of the other boundary divisors and we find that
as expected the correct GV-invariants are produced once we require the free energies to be regular at the orbifold point.
This way also all polynomial ambiguities $a_I$ are fixed uniquely. Here, in contrast to (\ref{M1}) we find that 
the $a_I$ are fixed without exception solely through the regularity constraint, no use of the fiber invariants is needed.

%%%%%%%%%%%%%%%%%%%%%%%%%%%%%%%%%%%%%%%%%%%%%%%%%%%%%%%%%%%%%%%%%%%%%%%%%%%%%%%%%%%%%%%%%%%%
\section{Conclusion} 

In this paper we solve the holomorphic anomaly equations of the $B$-model for 
two compact two parameter Calabi-Yau threefolds, which admit a regular $K3$ 
fibration with Picard number two.  The direct integration approach is  for the first 
time applied to compact Calabi-Yau spaces with more than one parameter. In the n-moduli case 
the $\frac{n(n+5)}{2}+1$ anholomorphic generators of the ring of modular functions, 
the propagators introduced by Berchadsky, Ceccoti, Ooguri and Vafa, and the derivative 
of the K\"ahler potential  have to obey compatibility conditions, which require the 
choice of holomorphic terms. An important feature is the closing  of the ring under 
the covariant derivatives, which appear on the righthand side of the holomorphic anomaly 
equation. This requires the determination of certain holomorphic terms, which 
for  the cases at hand are found in section \ref{directintegration}.  

The moduli space of these two parameter models is quite rich. In particular 
for type IIA compactifications on these manifolds the large $\mathbb{P}^1$ 
base limit corresponds to known heterotic compactifications, which are generically  weakly coupled in 
this limit. The one-loop calculation in the heterotic string for the 
graviton graviphoton interaction yields all genus informations for the type II string 
in the K3 fibre directions in terms of modular forms of weight $-\frac{3}{2}$ with 
various pole orders at the cusp. In section \ref{WeakCoupling} we present a 
systematic method using the Ranking- Cohen bracket to obtain these modular forms 
following work of Zagier. The formalism allows to construct these forms in general 
once the self intersection in the Picard lattice of the K3 and a fibration 
parameter are given. The all genus information in the fibre direction 
provides additional boundary conditions.

The heterotic string has near the weak coupling limit a $SU(2)$ gauge symmetry enhancement, 
which is nonperturbatively described by an embedding of the N=2 Seiberg-Witten gauge symmetry 
elliptic geometry in the B-model geometry. In particular near the point where 
the monopole becomes light, which is at a special point near the weak coupling divisor 
and the conifold divisor, we find that the gauge theory gap condition~\cite{HK}\cite{HKQ} is 
promoted to the compact Calabi-Yau space. The gap condition in the full model is stronger than in the one 
parameter case as for all negative powers of the normal coordinate to the conifold 
all orders in the transversal coordinate vanish. As a consequence this gap 
condition is strong enough to fix the two parameter ambiguity at the conifold 
completely.

Another important point is the Gepner point, which is in our models a 
$\mathbb{Z}_8$ and $\mathbb{Z}_{12}$ orbifold. We can consistently  
impose regularity at this point and obtain predictions for the rational 
orbifold Gromov-Witten invariants up to genus 4.

One of the most important divisors in the moduli space is the strong coupling divisor, 
which corresponds on the $A$-model side to vanishing size of the $\mathbb{P}^1$-base 
and hence to a strongly coupled heterotic string. We find for $g=2,3$ that the leading 
singularity is $t_{s,n}^{2g-2}$ followed by a gap structure. An explanation 
of the leading singularity can be given by the effect of the massless hyper - 
and vector multiplets in the effective action. However we found as in one 
parameter models with a more complicated massless spectrum that the gap structure can 
in principle vanish due non-trivial interactions between the light particles. 
This seems to be the case for $g=4$. If the gap structure was present at higher 
genus or if  these interactions could be understood more systematically  
the compact K3 fibrations would be completely integrable.        
         
\subsection*{Acknowledgments}

We gratefully acknowledge discussions with Emanuel Scheidegger and especially 
Don Zagier about the modular forms in the K3 fiber direction. We further like 
to thank Thomas Grimm, Tae-Won Ha, Denis Klevers, Marco Rauch, Piotr 
Sulkowski and Thomas Wotschke for useful comments and Denis Klevers for 
reading the manuscript. This work was supported in parts by the Hausdorff-Institut  
for Mathematics. B.H. is supported by the German Excellence Initiative via the BCGS.

\newpage

%%%%%%%%%%%%%%%%%%%%%%%%%%%%%%%%%%%%%%%%%%%%%%%%%%%%%%%%%%%%%%%%%%%%%%%%%%%%%%%%%%%%%%%%%%%
\appendix

\section{Propagator expansions}

Here we list the propagator expansions around important divisors of the moduli space.

For the model $M_1$ we obtain the following results.

\begin{itemize}

\item Conifold:

\begin{eqnarray}
  S^{1,1}_c & = & -\frac{11}{32} z_{c,1} - \frac{641}{1024} z_{c,1}^2 - \frac{4119}{16384} z_{c,1}^3
                  - \frac{z_{c,1}}{16 z_{c,2}} - \frac{47 z_{c,1}^2}{512 z_{c,2}}
                  - \frac{75 z_{c,1}^3}{4096 z_{c,2}}  \nonumber \\
  ~         & ~ & + \frac{5625 z_{c,1}^4}{1048576 z_{c,2}} + \frac{2591 z_{c,1}^2 z_{c,2}}{49152} 
                  + \cO(z_c^4), \nonumber \\
  S^{1,2}_c & = & \frac{47}{512} z_{c,1} z_{c,2} + \frac{1329 z_{c,1}^2 z_{c,2}}{16384} 
                  + \frac{2281}{24576} z_{c,1} z_{c,2}^2 + \cO(z_c^4) \nonumber \\
  S^{2,2}_c & = & \frac{z_{c,2}}{16} + \frac{29}{256} z_{c,2}^2 - \frac{3}{256} z_{c,1} z_{c,2}^2 
                  - \frac{601}{12288}z_{c,2}^3 + \cO(z_c^4), \nonumber \\
  S^1_c     & = & -\frac{25}{192} z_{c,1} z_{c,2} - \frac{25}{192} z_{c,1}^2 z_{c,2} + \frac{401}{2560} z_{c,1} z_{c,2}^2
                  + \cO(z_c^4), \nonumber \\
  S^2_c     & = & - \frac{z_{c,2}}{16} + \frac{25}{256} z_{c,2}^2 + \frac{25}{384} z_{c,1} z_{c,2}^2 
                  - \frac{1131}{8192} z_{c,2}^3 + \cO(z_c^4), \nonumber \\
  S_c       & = & \frac{z_{c,2}}{32} - \frac{79}{512} z_{c,2}^2 - \frac{91}{1536} z_{c,1} z_{c,2}^2
                  + \frac{12733}{24567} z_{c,2}^3 + \cO(z_c^4).
\end{eqnarray}

\item Strong coupling locus:

\begin{eqnarray}
  S^{1,1}_s & = & -\frac{1}{16} + \frac{z_{s,1}^2}{8} - \frac{z_{s,1}^4}{16} + \frac{9}{16384} z_{s,1}^2 z_{s,2}^2
                  + \cO(z_s^5), \nonumber \\
  S^{1,2}_s & = & -\frac{1}{16} z_{s,1} z_{s,2} + \frac{1}{16} z_{s,1}^3 z_{s,2} + \frac{19}{256} z_{s,1} z_{s,2}^2
                  - \frac{123}{32768} z_{s,1} z_{s,2}^3 + \cO(z_s^5), \nonumber \\
  S^{2,2}_s & = & -\frac{1}{16} z_{s,1}^2 z_{s,2}^2 - \frac{11}{128} z_{s,2}^3 + \frac{845}{32768} z_{s,2}^4 
                  + \cO(z_s^5), \nonumber \\
  S^1_s     & = & - \frac{3}{512} z_{s,1} z_{s,2} + \frac{3}{512} z_{s,1}^3 z_{s,2} - \frac{141}{65536} z_{s,1} z_{s,2}^2
                  + \cO(z_s^5), \nonumber \\
  S^2_s     & = & \frac{3}{256} z_{s,2}^2 - \frac{3}{512} z_{s,1}^2 z_{s,2}^2 + \frac{141}{32768} z_{s,2}^3
                  + \frac{3357}{1048576} z_{s,2}^4 + \cO(z_s^5), \nonumber \\
  S_s       & = & -\frac{27}{32768} z_{s,2}^2 + \frac{9}{16384} z_{s,1}^2 z_{s,2}^2 - \frac{2151}{2097152} z_{s,2}^3
                  \nonumber \\ 
            & ~ & - \frac{1572723}{1073741824} z_{s,2}^4 + \cO(z_s^5).
\end{eqnarray}
 
\item Gepner point:

\begin{eqnarray}
  S^{1,1}_o & = & \frac{5}{512} \frac{z_{o,2}}{z_{o,1}^2} + \frac{z_{o,2}^3}{6144 z_{o,1}^2}
                  + \frac{11}{3072} z_{o,1}^2 + \frac{z_{o,2}^5}{12288 z_{o,1}^2}
                  + \frac{z_{o,1}^2 z_{o,2}^2}{3072} + \cO(z_o^5), \nonumber \\
  S^{1,2}_o & = & -\frac{3}{128 z_{o,1}^3} + \frac{1}{768} z_{o,1} z_{o,2} + \frac{3}{128} \frac{z_{o,2}^2}{z_{o,1}^3}
                  + \cO(z_o^4), \nonumber \\                  
  S^{2,2}_o & = & \frac{5}{64} - \frac{5}{64} z_{o,2}^2 + \cO(z_o^5), \nonumber \\
  S^1_o     & = & -\frac{5}{2048} \frac{z_{o,2}}{z_{o,1}^3} - \frac{9}{32768} \frac{z_{o,2}^3}{z_{o,1}^3} 
                  - \frac{17}{1288} z_{o,1} - \frac{171}{1310720} \frac{z_{o,2}^5}{z_{o,1}^3}
                  + \cO(z_o^3), \nonumber \\
  S^2_o     & = & \frac{1}{128 z_{o,1}^4} -\frac{1}{128} \frac{z_{o,2}^2}{z_{o,4}^4}- \frac{z_{o,2}}{256}
                  + \frac{3}{4096} z_{o,2}^3 - \frac{11}{32256} z_{o,1}^4 + \cO(z_o^5), \nonumber \\
  S_o       & = & -\frac{1}{2048} \frac{z_{o,2}}{z_{o,1}^4} + \frac{3}{32768} \frac{z_{o,2}^3}{z_{o,1}^4}
                  + \frac{5}{12288} + \frac{57}{1310720} \frac{z_{o,2}^5}{z_{o,1}^4} + \frac{9}{65536} z_{o,2}^2 
                  \nonumber \\
  ~         & ~ & + \frac{1605}{58720256} \frac{z_{o,2}^7}{z_{o,1}^4} + \cO(z_o^4).
\end{eqnarray}
 
\end{itemize}

The model $M_2$ admits the following expansions.

\begin{itemize}

\item Conifold:

\begin{eqnarray}
  S^{1,1}_c & = & - \frac{z_{c,1}}{8 z_{c,2}} - \frac{23}{36} z_{c,1} - \frac{47}{256} \frac{z_{c,1}^2}{z_{c,2}} 
                  - \frac{1339}{1152} z_{c,1}^2 - \frac{75}{2048} \frac{z_{c,1}^3}{z_{c,2}} \nonumber \\
  ~         & ~ & + \frac{11195}{124416} z_{c,1}^2 z_{c,2} + \frac{5625}{524288} \frac{z_{c,1}^4}{z_{c,2}}
                  + \cO(z_c^4), \nonumber \\
  S^{1,2}_c & = & \frac{97}{576} z_{c,1} z_{c,2} + \frac{2719}{18432} z_{c,1}^2 z_{c,2} 
                  + \frac{10897}{62208} z_{c,1} z_{c,2}^2 + \cO(z_c^4), \nonumber \\
  S^{2,2}_c & = & \frac{z_{c,2}}{8} + \frac{67}{288} z_{c,2}^2 - \frac{5}{288} z_{c,1} z_{c,2}^2 
                  - \frac{2581}{31104} z_{c,2}^3 + \cO(z_c^4), \nonumber \\
  S^1_c     & = & - \frac{17}{72} z_{c,1} z_{c,2} - \frac{17}{72} z_{c,1}^2 z_{c,2} 
                  + \frac{1831}{6480} z_{c,1} z_{c,2}^2 + \cO(z_c^4), \nonumber \\
  S^2_c     & = & - \frac{z_{c,2}}{8} + \frac{17}{96} z_{c,2}^2 + \frac{17}{144} z_{c,1} z_{c,2}^2 
                  - \frac{1741}{6912} z_{c,2}^3 + \cO(z_c^4), \nonumber \\
  S_c       & = & \frac{z_{c,2}}{16} - \frac{169}{576} z_{c,2}^2 - \frac{7}{64} z_{c,1} z_{c,2}^2
                  + \frac{19999}{20736} z_{c,2}^3 + \cO(z_c^4).
\end{eqnarray}

\item Strong coupling locus:

\begin{eqnarray}
  S^{1,1}_s & = & - \frac{1}{8} + \frac{z_{s,1}^2}{4} - \frac{z_{s,1}^4}{8} + \frac{25}{41472} z_{s,1}^2 z_{s,2}^2 
                  + \cO(z_s^5), \nonumber \\
  S^{1,2}_s & = & -\frac{1}{8} z_{s,1} z_{s,2} + \frac{1}{8} z_{s,1}^3 z_{s,2} + \frac{41}{288} z_{s,1} z_{s,2}^2
                  - \frac{145}{27648} z_{s,1} z_{s,2}^3 + \cO(z_s^5), \nonumber \\
  S^{2,2}_s & = & -\frac{1}{8} z_{s,1}^2 z_{s,2}^2 - \frac{23}{144} z_{s,2}^3 + \frac{3797}{82944} z_{s,2}^4
                  + \cO(z_s^5), \nonumber \\
  S^1_s     & = & - \frac{5}{576} z_{s,1} z_{s,2} - \frac{485}{165888} z_{s,1} z_{s,2}^2 
                  + \cO(z_s^4), \nonumber \\
  S^2_s     & = & \frac{5}{288} z_{s,2}^2 - \frac{5}{576} z_{s,1}^2 z_{s,2}^2 + \frac{485}{82944} z_{s,2}^3
                  + \frac{26065}{5971968} z_{s,2}^4 + \cO(z_s^5), \nonumber \\
  S_s       & = & -\frac{25}{27648} z_{s,2}^2 - \frac{13225}{11943936} z_{s,2}^3 + \cO(z_s^4).
\end{eqnarray}

\item Gepner point:

\begin{eqnarray}
  S^{1,1}_o & = & \frac{13}{1728} \frac{z_{o,2}}{z_{o,1}^4} + \frac{z_{o,2}}{6912 z_{o,1}^4} 
                  + \frac{z_{o,2}^5}{13824 z_{o,1}^4} + \frac{19}{6912} z_{o,1}^2 
                  + \cO(z_o^3), \nonumber \\
  S^{1,2}_o & = & -\frac{1}{36 z_{o,1}^5} + \frac{1}{768} z_{o,1} z_{o,2} + \frac{z_{o,2}^2}{36 z_{o,1}^5}
                  + \cO(z_o^4), \nonumber \\
  S^{2,2}_o & = & \frac{7}{48} - \frac{7}{48} z_{o,2}^2 + \cO(z_O^4), \nonumber \\
  S^1_o     & = & - \frac{z_{o,2}}{648 z_{o,1}^5} - \frac{25}{139968} \frac{z_{o,2}^3}{z_{o,1}^5}
                  - \frac{205}{2519424} \frac{z_{o,2}^5}{z_{o,1}^5} - \frac{19}{20736} z_{o,1} \nonumber \\
  ~         & = & - \frac{190045}{3809369088} \frac{z_{o,2}^7}{z_{o,1}^5} + \cO(z_o^3), \nonumber \\
  S^2_o     & = & \frac{1}{144 z_{o,1}^6} - \frac{z_{o,2}}{144 z_{o,1}^6} - \frac{z_{o,2}}{288} 
                  + \frac{25 z_{o,2}^3}{31104} + \cO(z_o^3), \nonumber \\
  S_o       & = & -\frac{z_{o,2}}{1296 z_{o,1}^6} + \frac{25}{559872} \frac{z_{o,2}^3}{z_{o,1}^6}
                  + \frac{205}{10077696} \frac{z_{o,2}^5}{z_{o,1}^6} + \frac{11}{41472} \nonumber \\
  ~         & ~ & -\frac{85 z_{o,2}^2}{1492992} + \frac{190045}{15237476352} \frac{z_{o,2}^7}{z_{o,1}^6} + \cO(z_o^3).
\end{eqnarray}

\end{itemize}

\newpage

\section{Gopakumar-Vafa invariants}
\label{GVInvariants}

\begin{table}[!h]
\centering
\begin{tabular}[h]{|c|cccccccc|} 
\hline
      & $d_1$ & 0 & 1   & 2     &    3    &    4       &     5        & 6               \\
\hline                                                                                  
$d_2$ &       &   &     &       &         &            &              &                 \\
0     &       & 0 & 640 & 10032 & 288384  & 10979984   & 495269504    & 24945542832     \\
1     &       & 4 & 640 & 72224 & 7539200 & 757561520  & 74132328704  & 7117563990784   \\
2     &       & 0 & 0   & 10032 & 7539200 & 2346819520 & 520834042880 & 95728361673920  \\
3     &       & 0 & 0   & 0     & 288384  & 757561520  & 520834042880 & 212132862927264 \\
4     &       & 0 & 0   & 0     & 0       & 10979984   & 74132328704  & 95728361673920  \\
5     &       & 0 & 0   & 0     & 0       & 0          & 495269504    & 7117563990784   \\
6     &       & 0 & 0   & 0     & 0       & 0          & 0            & 24945542832     \\
\hline
\end{tabular}
\caption{Instanton numbers $n^{g=0}_{d_1 d_2}$ of $\P_4^{(1,1,2,2,2)}[8]$}
\end{table}

\begin{table}[!h]
\centering
\begin{tabular}[h]{|c|cccccccc|} 
\hline
      & $d_1$ & 0 & 1 & 2 &    3  &    4     &     5       & 6               \\
\hline                                                                                  
$d_2$ &       &   &   &   &       &          &             &                 \\
0     &       & 0 & 0 & 0 & -1280 & -317864  & -36571904   & -3478899872     \\
1     &       & 0 & 0 & 0 & 2560  & 1047280  & 224877056   & 36389051520     \\
2     &       & 0 & 0 & 0 & 2560  & 15948240 & 12229001216 & 4954131766464   \\
3     &       & 0 & 0 & 0 & -1280 & 1047280  & 12229001216 & 13714937870784  \\
4     &       & 0 & 0 & 0 & 0     & -317864  & 224877056   & 4954131766464   \\
5     &       & 0 & 0 & 0 & 0     & 0        & -36571904   & 36389051520     \\
6     &       & 0 & 0 & 0 & 0     & 0        & 0           & -3478899872     \\
\hline
\end{tabular}
\caption{Instanton numbers $n^{g=1}_{d_1 d_2}$ of $\P_4^{(1,1,2,2,2)}[8]$}
\end{table}

\begin{table}[!h]
\centering
\begin{tabular}[h]{|c|cccccccc|} 
\hline
      & $d_1$ & 0 & 1 & 2 & 3 & 4     & 5        & 6            \\
\hline                                                                                  
$d_2$ &       &   &   &   &   &       &          &              \\
0     &       & 0 & 0 & 0 & 0 & 472   & 875392   & 220466160    \\
1     &       & 0 & 0 & 0 & 0 & -1232 & -2540032 & -1005368448  \\
2     &       & 0 & 0 & 0 & 0 & 848   & 9699584  & 21816516384  \\
3     &       & 0 & 0 & 0 & 0 & -1232 & 9699584  & 132874256992 \\
4     &       & 0 & 0 & 0 & 0 & 472   & -2540032 & 21816516384  \\
5     &       & 0 & 0 & 0 & 0 & 0     & 875392   & -1005368448  \\
6     &       & 0 & 0 & 0 & 0 & 0     & 0        & 220466160    \\
\hline
\end{tabular}
\caption{Instanton numbers $n^{g=2}_{d_1 d_2}$ of $\P_4^{(1,1,2,2,2)}[8]$}
\end{table}

\begin{table}[!h]
\centering
\begin{tabular}[h]{|c|cccccccc|} 
\hline
      & $d_1$ & 0 & 1 & 2 & 3 & 4   & 5     & 6         \\
\hline                                                                                  
$d_2$ &       &   &   &   &   &     &       &           \\
0     &       & 0 & 0 & 0 & 0 & 8   & -2560 & -6385824  \\
1     &       & 0 & 0 & 0 & 0 & -24 & 3840  & 20133504  \\
2     &       & 0 & 0 & 0 & 0 & 24  & 2560  & -19124704 \\
3     &       & 0 & 0 & 0 & 0 & -24 & 2560  & 23433600  \\
4     &       & 0 & 0 & 0 & 0 & 8   & 3840  & -19124704 \\
5     &       & 0 & 0 & 0 & 0 & 0   & -2560 & 20133504  \\
6     &       & 0 & 0 & 0 & 0 & 0   & 0     & -6385824  \\
\hline
\end{tabular}
\caption{Instanton numbers $n^{g=3}_{d_1 d_2}$ of $\P_4^{(1,1,2,2,2)}[8]$}
\end{table}

\begin{table}[!h]
\centering
\begin{tabular}[h]{|c|ccccccccc|} 
\hline
      & $d_1$ & 0 & 1 & 2 & 3 & 4   & 5     & 6  &7         \\
\hline                                                                                  
$d_2$ &       &   &   &   &   &     &       &  &          \\
0     &       & 0 & 0 & 0 & 0 & 0  &  0 &  50160 &  101090432  \\
1     &       & 0 & 0 & 0 & 0 & 0 &   0  & -160512 & -355794944 \\
2     &       & 0 & 0 & 0 & 0 & 0  & 0  & 220704 &478526720\\
3     &       & 0 & 0 & 0 & 0 & 0 & 0  &   56160 & -366614784\\
4     &       & 0 & 0 & 0 & 0 & 0   & 0  &  220704 &-366614784\\
5     &       & 0 & 0 & 0 & 0 & 0   & 0 &  -160512&478526720\\
6     &       & 0 & 0 & 0 & 0 & 0   & 0    &  50160&  - \\
\hline
\end{tabular}
\caption{Instanton numbers $n^{g=4}_{d_1 d_2}$ of $\P_4^{(1,1,2,2,2)}[8]$}
\end{table}

\newpage

\begin{table}[!h]
\centering
\begin{tabular}[h]{|c|ccccccc|} 
\hline
      & $d_1$ & 0 & 1    & 2       & 3          & 4             & 5                \\
\hline                                                                                  
$d_2$ &       &   &      &         &            &               &                  \\
0     &       & 0 & 2496 & 223752  & 38637504   & 9100224984    & 2557481027520    \\
1     &       & 2 & 2496 & 1941264 & 1327392512 & 861202986072  & 540194037151104  \\
2     &       & 0 & 0    & 223752  & 1327392512 & 2859010142112 & 4247105405354496 \\
3     &       & 0 & 0    & 0       & 38637504   & 861202986072  & 4247105405354496 \\
4     &       & 0 & 0    & 0       & 0          & 9100224984    & 540194037151104  \\
5     &       & 0 & 0    & 0       & 0          & 0             & 2557481027520    \\
6     &       & 0 & 0    & 0       & 0          & 0             & 0                \\
\hline
\end{tabular}
\caption{Instanton numbers $n^{g=0}_{d_1 d_2}$ of $\P_4^{(1,1,2,2,6)}[12]$}
\end{table}

\begin{table}[!h]
\centering
\begin{tabular}[h]{|c|ccccccc|} 
\hline
      & $d_1$ & 0 & 1 & 2    & 3        & 4           & 5               \\
\hline                                                                                  
$d_2$ &       &   &   &      &          &             &                 \\
0     &       & 0 & 0 & -492 & -1465984 & -1042943520 & -595277880960   \\
1     &       & 0 & 0 & 480  & 2080000  & 3453856440  & 3900245149440   \\
2     &       & 0 & 0 & -492 & 2080000  & 74453838960 & 313232037949440 \\
3     &       & 0 & 0 & 0    & -1465984 & 3453856440  & 313232037949440 \\
4     &       & 0 & 0 & 0    & 0        & -1042943520 & 3900245149440   \\
5     &       & 0 & 0 & 0    & 0        & 0           & -595277880960   \\
\hline
\end{tabular}
\caption{Instanton numbers $n^{g=1}_{d_1 d_2}$ of $\P_4^{(1,1,2,2,6)}[12]$}
\end{table}

\begin{table}[!h]
\centering
\begin{tabular}[h]{|c|ccccccc|} 
\hline
      & $d_1$ & 0 & 1 & 2  & 3    & 4         & 5             \\
\hline                                                                                  
$d_2$ &       &   &   &    &      &           &               \\
0     &       & 0 & 0 & -6 & 7488 & 50181180  & 72485905344   \\
1     &       & 0 & 0 & 8  & 0    & -73048296 & -194629721856 \\
2     &       & 0 & 0 & -6 & 0    & 32635544  & 2083061531520 \\
3     &       & 0 & 0 & 0  & 7488 & -73048296 & 2083061531520 \\
4     &       & 0 & 0 & 0  & 0    & 50181180  & -194629721856 \\
5     &       & 0 & 0 & 0  & 0    & 0         & 72485905344   \\
\hline
\end{tabular}
\caption{Instanton numbers $n^{g=2}_{d_1 d_2}$ of $\P_4^{(1,1,2,2,6)}[12]$}
\end{table}

\begin{table}[!h]
\centering
\begin{tabular}[h]{|c|ccccccc|} 
\hline
      & $d_1$ & 0 & 1 & 2 & 3 & 4       & 5           \\
\hline                                                                                  
$d_2$ &       &   &   &   &   &         &             \\
0     &       & 0 & 0 & 0 & 0 & -902328 & -5359699200 \\
1     &       & 0 & 0 & 0 & 0 & 1357500 & 10139497472 \\
2     &       & 0 & 0 & 0 & 0 & -822968 & -7645673856 \\
3     &       & 0 & 0 & 0 & 0 & 1357500 & -7645673856 \\
4     &       & 0 & 0 & 0 & 0 & -902328 & 10139497472 \\
5     &       & 0 & 0 & 0 & 0 & 0       & -5359699200 \\
\hline
\end{tabular}
\caption{Instanton numbers $n^{g=3}_{d_1 d_2}$ of $\P_4^{(1,1,2,2,6)}[12]$}
\end{table}

\begin{table}[!h]
\centering
\begin{tabular}[h]{|c|ccccccc|} 
\hline
      & $d_1$ & 0 & 1 & 2 & 3 & 4       & 5           \\
\hline                                                                                  
$d_2$ &       &   &   &   &   &         &             \\
0     &       & 0 & 0 & 0 & 0 & 1164    & 228623232   \\
1     &       & 0 & 0 & 0 & 0 & -1820   & -376523648  \\
2     &       & 0 & 0 & 0 & 0 & 2768    & 144351104   \\
3     &       & 0 & 0 & 0 & 0 & -1820   & 144351104   \\
4     &       & 0 & 0 & 0 & 0 & 1164    & -376523648  \\
5     &       & 0 & 0 & 0 & 0 & 0       & 228623232   \\
\hline
\end{tabular}
\caption{Instanton numbers $n^{g=4}_{d_1 d_2}$ of $\P_4^{(1,1,2,2,6)}[12]$}
\end{table}

\newpage

\section{Verification of Gopakumar-Vafa invariants, by Sheldon Katz}

In this appendix, we compute several of the Gopakumar-Vafa invariants directly by the techniques of algebraic geometry, based on the principles in \cite{MtopII,KatzKlemmVafa}.

For each of the invariants $n^g_{d_1,d_2}$ that we consider, we will describe
the moduli space $\MM^g_{d_1,d_2}$ of connected curves of arithmetic genus $g$
and bidegree $(d_1,d_2)$.  In each case, $\MM^g_{d_1,d_2}$ is found to be
smooth, so by \cite{MtopII,KatzKlemmVafa} we expect that
\begin{equation}
n^g_{d_1,d_2}=(-1)^{\dim\MM^g_{d_1,d_2}}\ e\left(\MM^g_{d_1,d_2}\right),
\end{equation}
where $e\left(\MM^g_{d_1,d_2}\right)$ denotes the topological euler characteristic of $\MM^g_{d_1,d_2}$.  These computations can be used to both fix the
ambiguities and to provide an independent check.

It should be noted that while the calculations in this appendix provide all 
of the evidence needed for the purposes of this paper, they do not constitute
a rigorous mathematical proof, for several reasons.

First of all, while the Gopakumar-Vafa invariants can be given a rigorous
mathematical definition in terms of Gromov-Witten invariants $N^g_{d_1,d_2}$
by the identity in \cite{MtopII},

\begin{equation}
\label{eqn:gwgv}
Z=\mathrm{exp}\left(\sum_{d_1,d_2,g}N^g_{d_1,d_2}q_1^{d_1}q_2^{d_2}\lambda^{2g-2}\right)=
\sum_{d_1,d_2,g,m}n^g_{d_1,d_2}\frac1m\left(\sin\frac{m\lambda}2\right)^{2g-2}
q_1^{md_1}q_2^{md_2},
\end{equation}
there is not yet a direct a geometric definition of the $n^g_{d_1,d_2}$ that
can be used to even formulate a rigorous direct calculation.  There has been
some progress in this direction in recent years: there is a proposed direct
definition of $n^0_\beta$ for any Calabi-Yau threefold $X$ and class 
$\beta\in H^2(X,\Z)$ in terms of a moduli space of coherent sheaves 
\cite{katz}, with additional proposed definitions in \cite{joycesong} which
are conjectured to be equivalent definitions.  It is conjectured that with any 
of these definitions, the 
Aspinwall-Morrison formula 
\begin{equation}
N^0_\beta=\sum_{k\mid\beta}\frac{n^0_{\beta/k}}{k^3}
\end{equation}
holds.  The Aspinwall-Morrison formula
may be recognized as the coefficient of $\lambda^{-2}$ in (\ref{eqn:gwgv}).

It should also be remarked that the stable pair invariants of 
Pandharipande and Thomas \cite{pt} also conjecturally determine the 
topological string amplitudes, and the calculations in this appendix may
be interpreted as rigorous calculations of some of these PT 
invariants.  Even so,
infinitely many PT invariants are needed to compute a single GV invariant,
and we do not attempt to do that here.

\bigskip
We review some of the aspects of the geometry of $\P(1,1,2,2,2)[8]$.   The reader may want to compare 
with~\cite{CandelasMirrorII} where the geometry was also studied.

We first describe the toric variety $\P_\Sigma$ in which the Calabi-Yau is
embedded.  The variety $\P_\Sigma$ is a blowup of $\P(1,1,2,2,2)$.  From
(\ref{vectorsM1}) we see that the two $U(1)$ charge vectors are given by 
the rows of
\begin{equation}
\label{eqn:charges}
\left(
\begin{array}{cccccc}
0&0&1&1&1&1\\
1&1&0&0&0&-2
\end{array}
\right)
\end{equation}
In the geometry, these give the weights of a $\left(\C^*\right)^2$ action
on $\C^6$, as in (\ref{psigma}).
Letting $(x_1,\ldots,x_6)$ be coordinates on the $\C^6$, then the disallowed
set $Z$ of (2.1) is the union of the linear subspaces $x_1=x_2=0$ and
$x_3=x_4=x_5=x_6=0$.  

The bidegrees of the hypersurfaces $x_i=0$ of $\P_\Sigma$ may be read off
as the respective columns of (\ref{eqn:charges}).  Since $x_3,x_4,x_5$
have identical degrees, they are sections of a common line bundle, which we
call $L_1$.  Similarly $x_1,x_2$ are sections of a line bundle $L_2$.  For 
later use, we note that $L_1$ has six independent sections, as
$x_6x_1^2,\ x_6x_1x_2$, and $x_6x_2^2$ are also sections of $L_1$.

The Calabi-Yau hypersurface $M$ is obtained as the zero locus of a 
homogeneous polynomial of bidegree $(4,0)$, whose equation is
of the form
\begin{equation}
\label{eqn:cyeqn}
\sum_{i=0}^4x_6^if_{4-i}(x_3,x_4,x_5)g_{2i}(x_1,x_2)=0,
\end{equation} 
where the subscripts on $f$ and $g$ indicate the degree of a corresponding
homogeneous polynomial in the indicated variables.

We will also consider the $L_i$ as line bundles on $M$ by restriction.  

Projection to the first two coordinates defines a
projection $\pi:M\to\P^1$, whose fibers can be identified (noncanonically)
with a degree~4 K3 
hypersurface in a $\P^3$ with homogeneous coordinates $(x_3,x_4,x_5,x_6)$.
To see this, we fix a value of $(x_1,x_2)$ to specify the fiber.  Picking
a nonzero coordinate $x_1$ or $x_2$, we can use the second $\C^*$ to fix
its value to 1.  The first $\C^*$ then acts on $(x_3,x_4,x_5,x_6)$ as scalar
multiplication, so a $\P^3$ is obtained.  Then the equation (\ref{eqn:cyeqn})
is recognized as being homogeneous of degree~4 in the variables 
$(x_3,x_4,x_5,x_6)$.  

The hypersurface given by $x_6=0$ is the exceptional divisor of the blowup
of $\P(1,1,2,2,2)$.  From (\ref{eqn:cyeqn}) we see after putting $x_6=0$
that the exceptional divisor is a ruled surface, a family of $\P^1$s
parametrized by a plane curve $D$ with equation $f_4(x_3,x_4,x_5)=0$.  The 
curve $D$ has genus~3.

We are now prepared to compute some GV invariants.  Given a curve $C$ of
bidegree $(d_1,d_2)$, this means that
\begin{equation}
C\cdot L_1=d_1,\qquad C\cdot L_2=d_2.
\end{equation}

Let's first compute $n^3_{4,0}$.  Since $C\cdot L_2$ is the degree of
the projection of $C$ to $\P^1$ and $C\cdot L_2=0$, we conclude that
$C$ must be contained in a K3 fiber.  Let $\phi:\MM^3_{4,0}\to\P^1$ send
$C$ to the point of $\P^1$ corresponding to the fiber of $\pi$ that $C$
is contained in.  To find $e(\MM^3_{4,0})$, we need only describe the fibers
of $\phi$, i.e.\ the curves contained in a single K3 surface, and multiply
its euler characteristic with that of $\P^1$.

By the explicit description above,
we see that $C$ is identified as a curve of degree~4 in a degree~4 K3 surface
$S\subset \P^3$. By Castelnuovo theory~\cite{harris}, degree~4 genus~3 curves must be
contained in a hyperplane $H$.  Since $H\cap S$ has degree~4 and contains $C$,
we conclude that $C=H\cap S$.  So the fibers
of $\phi$ are identified with moduli space of hyperplanes in $\P^3$, itself
isomorphic to $\P^3$.  Since $\dim(\MM^3_{4,0})=4$ is even, we conclude
that
\begin{equation}
n^3_{4,0}=e(\P^1)e(\P^3)=8,
\end{equation} 
in agreement with (A.4).

\bigskip
We next calculate $n^3_{4,1}$.  Let $C\subset M$ be a curve of genus~3 with 
$C\cdot L_1=4$ and $C\cdot L_2=1$.  The last equality tells us that the
restricted projection $\pi|_C:C\to \P^1$ has degree~1.  Since there are no
degree~1 maps from irreducible curves of positive genus to $\P^1$, we conclude
that $C$ must be reducible, and in particular $C$ is of the form
$C=C_0\cup C_1$, where $C_0$
has genus~0 and is mapped isomorphically to $\P^1$ via $\pi$.  The curve
$C_0$ therefore has bidegree $(a,1)$ for some $a$ between 0 and 4.  The curve
$C_1$ therefore has bidegree $(4-a,0)$.
A priori,
$C_1$ need not be connected, but each connected 
component of $C_1$ must be contained 
in some K3 fiber $S$.

From $C_0\cdot L_2=1$ we see that $C_0$ meets each K3 fiber once, so
can only meet each connected component of $C_1$ at most once.  But
it is not possible within the bounds of Castelnuovo theory to distribute
a genus of 3 and degree of $4-a$ among different components of $C_1$,
unless it is a single connected curve in one K3 fiber and $a=0$ so that
$C_1$ has bidegree $(4,0)$.  As before, we
conclude that $C_1$ can be identified with the intersection of a K3
fiber $S$ with a hyperplane $H\subset\P^3$.
The curves
$C_0$ then have bidegree $(0,1)$ and
are precisely the rulings of the exceptional divisor parametrized by
the genus~3 curve $D$ described above.

We can now describe $\MM^3_{4,1}$.  Associating to $C$ the point of $D$ 
parametrizing $C_0$ and the point of $\P^1$ describing the fiber of $\pi$
which contains $C_1$, we get a map
\begin{equation}
\psi:\MM^3_{4,1}\to D\times\P^1.
\end{equation}

The fibers of $\psi$ correspond to hyperplanes $H\subset \P^3$ containing the
unique point of the $\P^3$ fiber (in the original toric variety) contained in
$C_0$.  This space is isomorphic to $\P^2$, as one linear condition has been
imposed on the $\P^3$ of all hyperplanes in $\P^3$. Since $\dim\MM^3_{4,1}=4$
is even, we conclude that
\begin{equation}
n^3_{4,1}=e(D\times \P^1)e(\P^2)=(2-2\cdot 3)(2)(3)=-24,
\end{equation}
in agreement with (A.4).

\bigskip\noindent
We need another technique for $n^4_{6,0}$.  As before, we conclude from $C\cdot L_2=0$ that
$C$ is contained in a K3 fiber of $\pi$ and thus can be identified with
a degree~6 genus~4 curve in a degree~4 K3 surface $S\subset\P^3$.

We assert that $C$ is also contained in a degree~2 hypersurface $Q\subset\P^3$.
To see this, we consider the restriction of degree~2 homogeneous polynomials
from $\P^3$ to $C$, i.e.\
\begin{equation}
r:H^0(\cO_{\P^3}(2))\to H^0(\cO_C(2)).
\end{equation}
Computing dimensions, we have $h^0(\cO_{\P^3}(2))=10$ and $h^0(\cO_C(2))
=2\cdot6+1-4=9$ by Riemann-Roch, so $r$ has a nontrivial kernel, a nonzero
degree~2 polynomial whose vanishing defines $Q$ as claimed.

Therefore $C\subset Q\cap S$.  But $Q\cap S$ is a curve of degree
$2\times 4=8$ containing $C$, so that $Q\cap S = C\cup C'$ for some curve
$C'$ of degree~2.  In other words, $C$ and $C'$ are related by liaison~\cite{ps}.

In general, if $C$ and $C'$ are related by liaison, i.e.\ $C\cup C'=H_e\cap H_f$
for hypersurface $H_e$ and $H_f$ of respective degrees $e$ and $f$, then
their genera $g$ and $g'$ are related by
\begin{equation}
g-g'=\frac12\left(e+f-4\right)\left(d-d'\right).
\end{equation}
In our situation, we learn that $C'$ has genus~0.  Putting everything we
know together, we see that $C'$ contributes to $n^0_{2,0}$.

We can now reverse the liaison construction to describe $\MM^4_{6,0}$.
By $n^0_{2,0}=10032$, we expect 10032 curves of the type $C'$ above.  Such
a curve lies in a K3 fiber which we identify as a degree~4
hypersurface $S\subset\P^3$.  Fixing one such curve $C_i'$, then any
degree~2 hypersurface $Q\subset\P^3$ containing $C'$ produces a
degree~6 genus~4 curve $C$ by liaison, i.e.\ $S\cap Q=C\cup C_i'$.  The
set of all such $C$ forms a component $\MM^{4,i}_{6,0}$ of
$\MM^4_{6,0}$.  We now describe $\MM^{4,i}_{6,0}$.

We study the kernel of the restriction map
\begin{equation}
s:H^0(\cO_{\P^3}(2))\to H^0(\cO_{C_i'}(2)).
\end{equation}
As before, $h^0(\cO_{\P^3}(2))=10$, and by Riemann-Roch $h^0(\cO_{C_i'}(2))=
2\cdot2+1-0=5$.  Furthermore, $s$ is surjective, either by regularity theory
or by direct calculation.  So $s$ has a 5-dimensional kernel, which gets
projectived in forming the moduli space.  Thus $\MM^{4,i}_{6,0}\simeq\P^4$.
We conclude that
\begin{equation}
n^4_{6,0}=10032 \ e(\P^4)=50160,
\end{equation}
in agreement with (A.5).

We finally turn to $n^4_{6,1}$, combining the liaison argument with the method 
for going from $n^3_{4,0}$ to $n^3_{4,1}$ above.  We start by noting
that any curve $C$ of genus~4 and bidegree $(6,1)$ is necessarily a union
$C=C_0\cup C_1$, where $C_0$ has bidegree $(0,1)$ and genus~0, and
$C_1$ has bidegree $(6,0)$ and genus~4, while $C_0$ and $C_1$ intersect
at a single point $p$.

As above, we have 10032 isomorphic components of this moduli space corresponding
to the choice of $C'$ used to construct $C_1$, and we describe
one of them.  The curve $C_0$ is parametrized by $D$ as in the case of $n^3_{4,1}$.  Once $C_0$ is fixed, the point $p$ is determined as the point where $C_0$
meets the K3 fiber containing $C'$ (here we are using $C'\cdot L_2=1$).
Then $C_1$ is parametrized by the family of degree~2 hypersurfaces $Q$ which
contain both $C'$ and $p$, and this is a $\P^3$.  In other words, the 
component of the moduli space is a $\P^3$ fibration over $D$.  Putting this
all together, we get $n^4_{6,1}=10032(2-2\cdot 3)(4) = -160512$,
in agreement with (A.5).

\bigskip
We will be more brief in the case of $\P(1,1,2,2,6)[12]$ since the ideas
and techniques are similar.  In this case, from the charges we see immediately
that the equation of the Calabi-Yau
hypersurface $M$ is of the form
\begin{equation}
\label{eqn:cy2eqn}
\sum_{i=0}^4x_6^if_{6-i}(x_3,x_4,x_5)g_{2i}(x_1,x_2)=0,
\end{equation}
where now $x_5$ is assigned a weight of~3 in determining the degree
of $f_{6-i}(x_3,x_4,x_5)$.  Again, the first two coordinates $(x_1,x_2)$
define a line bundle $L_1$ and a projection $\pi:M\to\P^1$.  The coordinates
$(x_3,x_4)$ are sections of a line bundle $L_2$, but now 
we see that $x_5$ is a section of $L_2^{\otimes3}$.  The exceptional divisor
is a ruled surface parametrized by the curve $f_6(x_3,x_4,x_5)=0$ in
$\P(1,1,3)$.  The curve $D$ has genus~2.

The fibers can be (noncanonically) identified with weighted K3
hypersurfaces $\P(1,1,3,1)[6]$ with coordinates $(x_3,x_4,x_5,x_6)$ by
the same method as before.  It is useful to project these K3
hypersurfaces $S$ onto the $\P^2$ with coordinates $(x_3,x_4,x_6)$ and
then from (\ref{eqn:cy2eqn}) we recognize that $S$ can be described as
the double cover  $\rho:S\to \P^2$ branched over a degree~6 plane curve $F$.
Curves $C$ with $C\cdot L_2=0$ are
necessarily contained in a K3 fiber, and then $C\cdot L_1$ coincides
with the degree of $\rho(C)$ as a curve in $\P^2$.  Note that some components
$C'$ of $\rho(C)$ may need to be counted with a multiplicity of~2 if 
$\rho:C\to \P^2$ is generically 2-1 over $C'$.

We now compute $n^2_{2,0}$.  From $C\cdot L_2=0$, we see that $C$ is
contained in a K3 fiber, giving a map $\phi:\MM^2_{2,0}\to \P^1$.  The
fibers of $\phi$ are identified with genus~2 curves $C\subset S$ with
$\rho(C)$ of degree~2 including multiplicity.  So $C$ is either isomorphic to a
degree~2 plane curve or is a double cover of a line $L\subset\P^1$,
branched over the 6~points $F\cap L$.  In the former case, $C$ would have genus
0, a contradiction, and in the second case $C$ has genus~2 by the 
Riemann-Hurwitz formula, which is exactly
what we need.  This identifies the fibers of $\phi$ with the set of lines
$L\subset \P^2$, which is itself a $\P^2$.  Thus $\MM^2_{2,0}$ has dimension~3 and
\begin{equation}
n^2_{2,0}=-e(\P^1)e(\P^2) = -6,
\end{equation}
in agreement with (A.8).

\medskip
The calculation of $n^2_{2,1}$ follows from that of $n^2_{2,0}$ in the same
way that $n^3_{4,1}$ followed from $n^3_{4,0}$ for $\P(1,1,2,2,2)[8]$.  We 
conclude that $C=C_0\cup C_1$, where $C_0$ has bidegree $(0,1)$ and is
parametrized by $D$, and $C_1$ is connected and of bidegree~$(2,0)$, and of
genus~2.

This description gives a map
\begin{equation}
\psi:\MM^2_{2,1}\to D\times\P^1
\end{equation}
with fibers the set of lines $L\subset \P^2$ passing through a fixed point
of $\P^2$.  This set is isomorphic to $\P^1$, and we conclude that
\begin{equation}
n^2_{2,1}=-e(D\times\P^1)e(\P^1)=-(2-2\cdot2)(2)(2)=8,
\end{equation}
in agreement with (A.8).

\medskip The last check will be the calculation of $n^2_{3,0}$.  We
can do this by liaison, but it will be more convenient to describe
the geometry directly.  

These arithmetic genus~2 curves $C$ lie on some K3
fiber and $\rho(C)$ has degree~3.  Note that $C$ cannot project
isomorphically onto a degree~3 plane curve $\rho(C)$ since then $C$
would have genus~1.  Therefore $C$ is reducible, and each component
either maps isomorphically to a degree~1 or degree~2 plane curve, or
else maps 2 to 1 onto a line.  The first two types of curves have
genus~0 and we have just seen that the last class of curves has
genus~2.  

The only possibility is that $C=C_1\cup C_2$ with $C_1$ projecting
isomorphically onto a line via $\rho$ and $C_2$ projecting 2-1 onto a 
line (not necessarily the same line).  Note that in this case $C_1\cap C_2$
is a single point, so $C$ indeed has arithmetic genus~2.

There are $n^0_{1,0}=2496$ such components $C_1$.  Once $C_1$ is fixed, then
the K3 fiber is fixed. As we have seen in the calculation of $n^2_{2,0}$, the
curve $C_2$ is parametrized by a $\P^2$.  Thus
\begin{equation}
n^2_{3,0}=2496\,e(\P^2)=7488,
\end{equation}
in agreement with (A.8).

We would like to verify $n^2_{3,1}=0$.  However, $\MM^2_{3,1}$ is singular,
so the simple techniques used above do not apply in this case.

For each of the 2496~genus~0 curves $C'$ of bidegree $(1,0)$, we will
describe three intersecting components of $\MM^2_{3,1}$, exhibiting the
claimed singularity.

As we have seen, each such curve $C'$ intersects a unique curve $C_1$ of
bidegree $(0,1)$.  Fix any K3 fiber, which necessarily meets $C_1$ in a 
point $p$, and let $C_2$ a curve of genus~2 and bidegree~$(2,0)$ in that
K3 fiber containing $p$.  
Then the curve $C=C'\cup C_1\cup C_2$ is the desired
curve.  This component is a $\P^1$-bundle over $\P^1$.

The second component of $\MM^2_{3,1}$ parametrizes curves $C=C'\cup C_1\cup
C_2$, where $C'$ and $C_1$ are as above, but now $C_2$ is any genus 2 curve
of bidegree $(2,0)$ in the same K3 fiber as $C'$.  We no longer require $C_2$
to meet $C_1$, since $C_2$ automatically meets $C'$.  This component is
a $P^2$.

To describe the third component, take any curve $C_1$ of
bidegree~$(0,1)$ and any curve $C_2$ meeting $C_1$ which has genus 2
and bidegree~$(2,0)$ in the same K3 fiber as $C'$.
Then the curve $C=C'\cup C_1\cup C_2$ is the desired curve.  This
component is a $\P^1$-bundle over $D$.

The three components intersect in the curves $C=C'\cup C_1\cup C_2$ as in the
previous paragraph, where now $C_1$ and $C'$ are constrained to
intersect.  This singular
locus is isomorphic to $\P^1$.

%%%%%%%%%%%%%%%%%%%%%%%%%%%%%%%%%%%%%%%%%%%%%%%%%%%%%%%%%%%%%%%%%%%%%%%%%%%%%%%%%%%%%%%%%%%%%

\end{document}